\documentclass[%
 amsmath,amssymb,
	aps,
	twocolumn,
	altaffilletter,
	nolongbibliography,
	numerical,
	flushbottom,
	secnumarabic,
	pra,
	superscriptaddress,
	floatfix,
	10pt
]{revtex4-2}

\usepackage[normalem]{ulem}
\usepackage{graphicx}% Include figure files
\usepackage{dcolumn}% Align table columns on decimal point
\usepackage{bm}
\usepackage{amsmath}% bold math
\usepackage{siunitx}
\usepackage{lipsum}
\usepackage{bbold}
\usepackage{hyperref}% add hypertext capabilities
\hypersetup{%
	colorlinks=true, pdfstartpage=1, pdfstartview=FitH, pdfborder={0 0 0},%
	breaklinks=true, pdfpagemode=UseNone, pageanchor=true, pdfpagemode=UseOutlines,%
	plainpages=false, bookmarksnumbered, bookmarksopen=true, bookmarksopenlevel=1,%
	hypertexnames=true, pdfhighlight=/O,
	urlcolor=blue, linkcolor=blue, citecolor=blue
}

\newcommand{\nspin}{N}

\newcommand{\ham}{H}

\newcommand{\hamz}{\ham_z}

\newcommand{\tf}{t_f}

\newcommand{\addition}[1]{#1}
\newcommand{\removal}[1]{}

\begin{document}

\preprint{APS/123-QED}

\title{Deep learning optimal quantum annealing schedules for random Ising models}% Force line breaks with \\

\author{Pratibha Raghupati Hegde}
\affiliation{Dipartimento di Fisica ``E. Pancini'', Universit\`a degli Studi di Napoli ``Federico II'', Complesso Universitario M. S. Angelo, via Cintia 21, 80126, Napoli, Italy}

\author{Gianluca Passarelli}
\affiliation{CNR-SPIN, c/o Complesso Universitario M. S. Angelo, via Cintia 21, 80126, Napoli, Italy}

\author{Giovanni Cantele}
\affiliation{CNR-SPIN, c/o Complesso Universitario M. S. Angelo, via Cintia 21, 80126, Napoli, Italy}

\author{Procolo Lucignano}
\affiliation{Dipartimento di Fisica ``E. Pancini'', Universit\`a degli Studi di Napoli ``Federico II'', Complesso Universitario M. S. Angelo, via Cintia 21, 80126, Napoli, Italy}

%\affiliation{%


%\affiliation{
%\affiliation{
%\affiliation{%



\begin{abstract}
A crucial step in the race towards quantum advantage is optimizing quantum annealing using ad-hoc annealing schedules. Motivated by recent progress in the field, we propose to employ long-short term memory (LSTM) neural networks to automate the search for optimal annealing schedules for random Ising models on regular graphs. By training our network using locally-adiabatic annealing paths, we are able to predict optimal annealing schedules for unseen instances and even larger graphs than those used for training. 

\end{abstract}

%Use showkeys class option if keyword
\maketitle

%\tableofcontents

\section{\label{sec:level1}Introduction}
%Quantum annealing is a method for solving hard optimization problems. The advancements in the quantum annealer hardware by D-Wave has made quantum annealing a promising technique for solving real world optimization problems. 
Quantum annealing (QA) is a heuristic technique to solve optimization problems by encoding their solutions in the ground states of appropriate spin Hamiltonians~\cite{kadowaki-nishimori:qa, Farhi:00}.
In the last years it has been applied to a vast variety of problems, ranging from physics, to biology, economy and more~\cite{Brooke1999,santoro:qa,PhysRevE.70.057701,PhysRevE.71.066707,Matsuda:2009uq,Perdomo-Ortiz2012,Ramsey-expt,ronnow:speedup,Rieffel:2015aa,Azinovic:2016uq,Mott:2017aa,Li:comp-bio-2017,Mandra:2017ab,Jiang2018,Venturelli2019,Smelyanskiy:2018aa,Zlokapa:2019ab}.  Extensive reviews on the subject can be found in Refs.~\cite{Tanaka:book,albash:aqc-review,Hauke:2019aa}.

The main idea is to initialize a quantum spin system in an easy-to-prepare ground state of a simple noninteracting  Hamiltonian $H_x$ and adiabatically deform it into a final Hamiltonian $H_z$, whose ground state encodes the solution of the optimization problem at hand. 
The time dependent driving protocol  is  $H(s)= A(s) H_x + B(s) H_z$, where $A(s)$ and $B(s)$ are the so called  annealing schedule functions, which satisfy the boundary conditions $A(0) =B(1)=1$ and $A(1)=B(0)=0$. A simple, traditional, approach to this parametrization is to choose a linear schedule $ s=\frac{t}{\tf} $, where $\tf$ is the annealing time, $A(s)=1-s$ and $B(s)=s$. The evolution to the ground state is guaranteed by the adiabatic theorem of quantum mechanics~\cite{Born1928}, when the system is evolved at a rate which is directly proportional to powers of the minimum energy gap between the ground state and the excited states~\cite{lidar:adiabatic-theorem, albash:aqc-review,kato:adiabatic-theorem, jansen:adiabatic-theorem}. 
%A rigorous treatment of the theorem also shows that the rate of evolution can scale proportional up to cube of the gaps~\cite{jansen:adiabatic-theorem}.} 
In general,  hard optimization problems are characterized by small energy gaps in the adiabatic evolution, therefore making the timescale to reach the ground state of the problem Hamiltonian very large~\cite{jorg:energy-gaps,lucas:np-complete}. That makes it impossible to attack these problems on real-world  Noisy Intermediate-Scale Quantum devices~\cite{Preskill2018}, in which decoherence and relaxation mechanisms set the upper limit for the time available for the (quantum) computation. In these devices new time scales  enter the game, together with the adiabatic time scale, due to the system-bath coupling~\cite{albash:decoherence,amin:speedup}. The dynamics is no longer unitary and some new ideas can be pursued to improve the quantum annealing performances~\cite{gpassarelli:qa4} such as pausing the dynamics~\cite{gpassarelli:qa1, marshall:pausing, marshal:pausing2, lidar:pausing} or applying reverse annealing schemes~\cite{gpassarelli:qa2,  gpassarelli:qa6, ohkuwa:reverse, lidar:reverse, venturelli:reverse}. %QUI AGGIUNGERE ANCHE REFERENZE DI ALTRI. 
In this manuscript, however, we focus on unitary dynamics and ignore the role of dissipation.  Also in this case, different strategies have been proposed to speed up QA, such as approximating counterdiabatic  terms using variational approaches~\cite{claeys:variational-sta, delcampo:sta, guery:sta, sels:sta, gpassarelli:qa3, TORRONTEGUI:sta, gpassarelli:qa5, crosson:sta, seki:sta}, using diabatic transitions~\cite{Somma:2012kx,brady_optimal_2021,venuti2021optimal,Crosson2021, farhi:sta,cote-diabatic}, or using catalyst Hamiltonians~\cite{durkin:catalystH,albash:catalystH}.

\begin{figure}[tb]
    \centering
    \includegraphics[width=\linewidth]{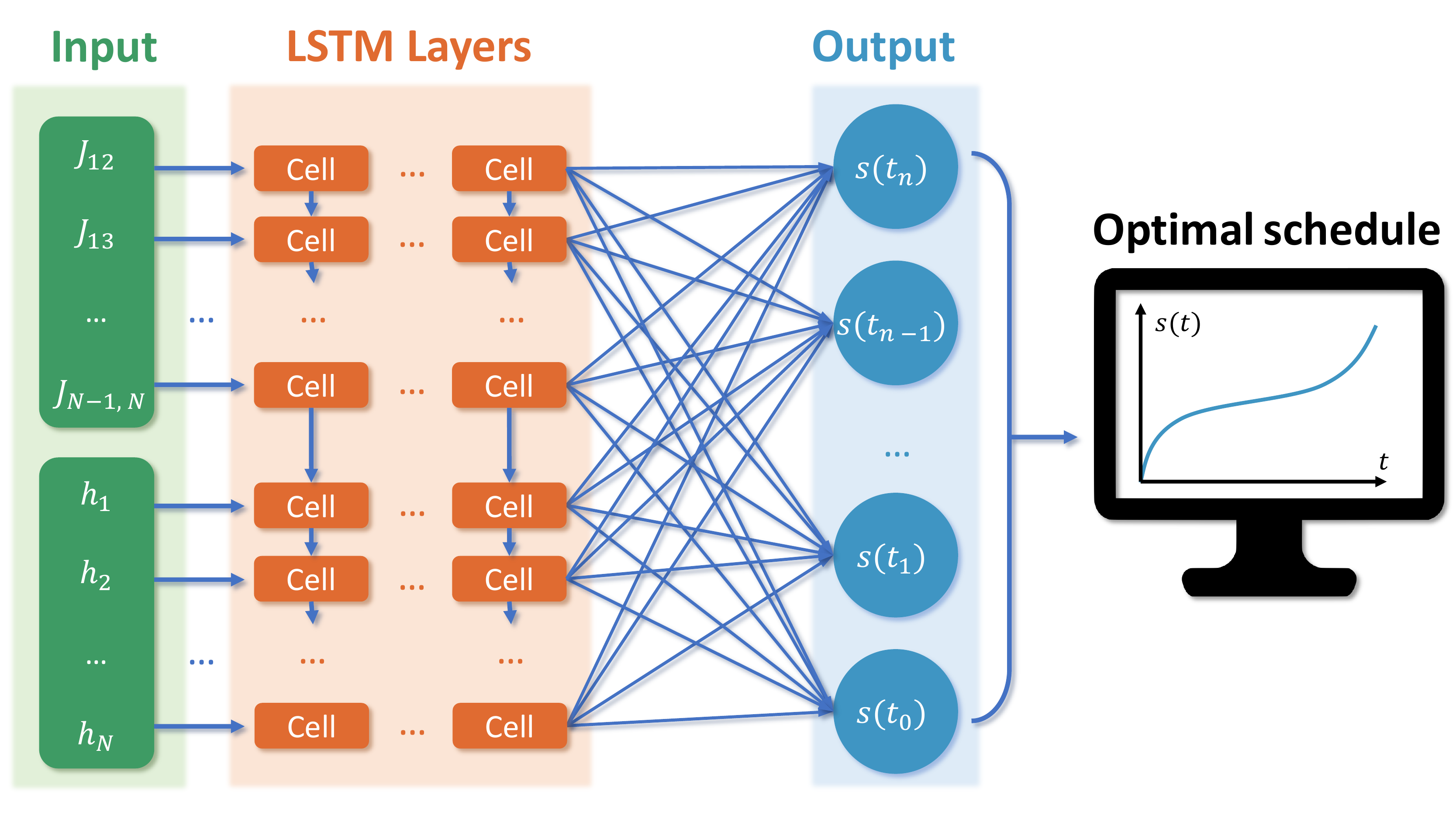}
    \caption{Schematic representation of using Long-Short Term Memory (LSTM) neural network for the prediction of optimal annealing schedules for random Ising models. The neural network is trained with the interactions and external longitudinal magnetic fields as input and the corresponding optimal schedules computed with Roland and Cerf protocol as output. The neural network contains several layers of LSTM cells and a final dense layer whose values represent the optimal schedule values $s(t)$ computed at time intervals between initial time and final annealing time. }
    \label{fig:lstm}
\end{figure}

One of the main routes to tackle this problem, in the unitary limit, is to engineer the annealing schedules as opposed to using the simple linear schedules. There have been many approaches to construct optimal schedules for the given problem in the recent past~\cite{local-adiab:roland-cerf,  khezri:qa-sch, morita:qa-sch, matsuura-timesch-optimization, susa-nishimori-timesch-optimization, chen:sch-ml-3sat}, and some of them~\cite{phegde:ga} are based on genetic algorithms, that have been also fruitfully applied to the embedding problem~\cite{choi_minor-embedding_2008, acampora:genetic,acampora:genetic2, zaman:pyqubo, Hen:embedding} for QA.

Since most of the  combinatorial optimization problems can be represented in terms of random Ising models~\cite{lucas:np-complete, barahona:isingmodel}, controlling and understanding the best annealing schedule for random Ising models is mandatory. However, constructing optimal annealing schedules for every random Ising model, using heuristic methods is a hard task, and recently machine learning based algorithms have been employed to this aim. 
For instance, in Ref.~\cite{chen:sch-ml-3sat}  the authors have used QZero machine learning algorithm~\cite{mitchell_machine_1997} to obtain optimal annealing schedules for 3-SAT problems. 

In this paper, we aim at training a suitable neural network with optimal schedules for a set of instances of random Ising Hamiltonians, with the goal to predict the optimal schedules for new test instances. In the recent work by N.~Mohseni \textit{et al.}~\cite{lstm-sch:mohseni} the authors have shown that Long Short Term Memory (LSTM) neural networks are efficient in learning the energy gaps in the adiabatic evolution of random Ising Hamiltonians. In~\cite{mohseni:lstm2}, authors have employed LSTM neural networks to learn the dynamics of many-body systems under random driving. We choose LSTM neural networks to learn the optimal schedules of random Ising models.

To train the LSTM neural networks,  we generate optimal annealing schedules for a given random Ising problem using local adiabaticity condition that has been, first, used by Roland and Cerf~\cite{local-adiab:roland-cerf} to obtain an optimal schedule for the adiabatic search problem (equivalent to Grover's search algorithm~\cite{grover}). The adiabatic search algorithm is to date, one of the very few examples in which a measurable (quadratic) quantum speedup has been shown in adiabatic quantum computation. Since then, this approach has been used several times in the literature for the optimal control of quantum annealing~\cite{mbeng:roland-cerf, Stefanatos:roland-cerf, Delvecchio:roland-cerf}.

This paper is organized as follows. In Sec.~\ref{sec:problem definition}, we present the random Ising model and we describe the  local adiabaticity condition and its application  to our problem. In Sec.~\ref{sec:neural-network} we describe the LSTM neural networks and the scheme of training. In Sec.~\ref{sec:results} we present the results of training performance of neural networks and the results of quantum annealing simulated using optimal schedules predicted from the neural network. Generating large enough training sets for machine learning is daunting because one has to integrate the exact local adiabatic condition multiple times for many random realizations of the same model, which becomes prohibitive as the system size grows. However, in Sec.~\ref{sec:extrapolation}, we show that with the LSTM architecture it is possible to train over locally adiabatic schedule of small systems and then generalize to larger systems making training easier. In Sec.~\ref{sec:permutation}, we test the learnability of symmetric properties of the Ising models by LSTM neural networks. In Sec.~\ref{sec:Conclusions} we discuss the final conclusions.

\begin{figure*}[tb]
    \centering
    \includegraphics[width=0.3\textwidth]{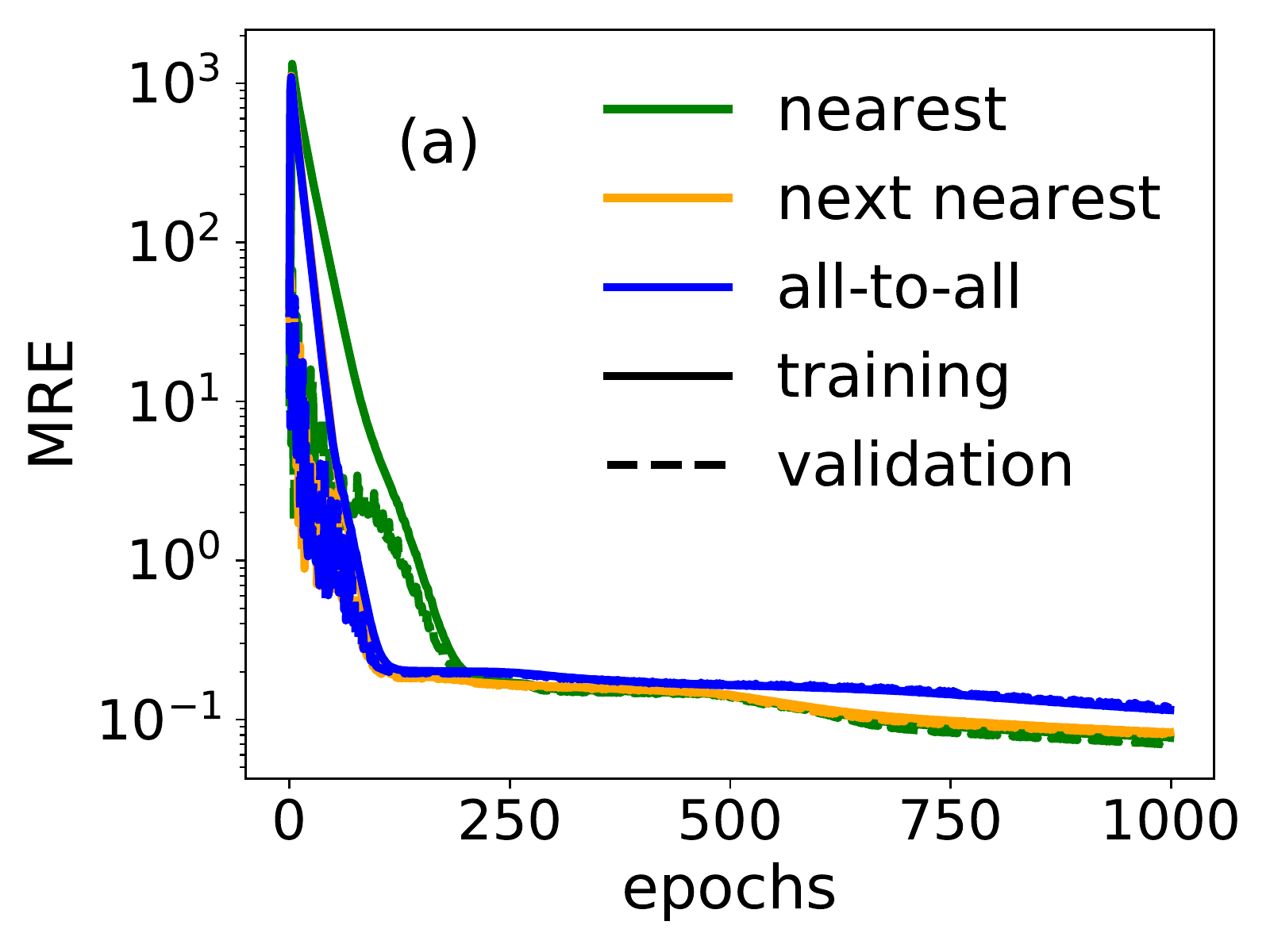}
    \includegraphics[width=0.3\textwidth]{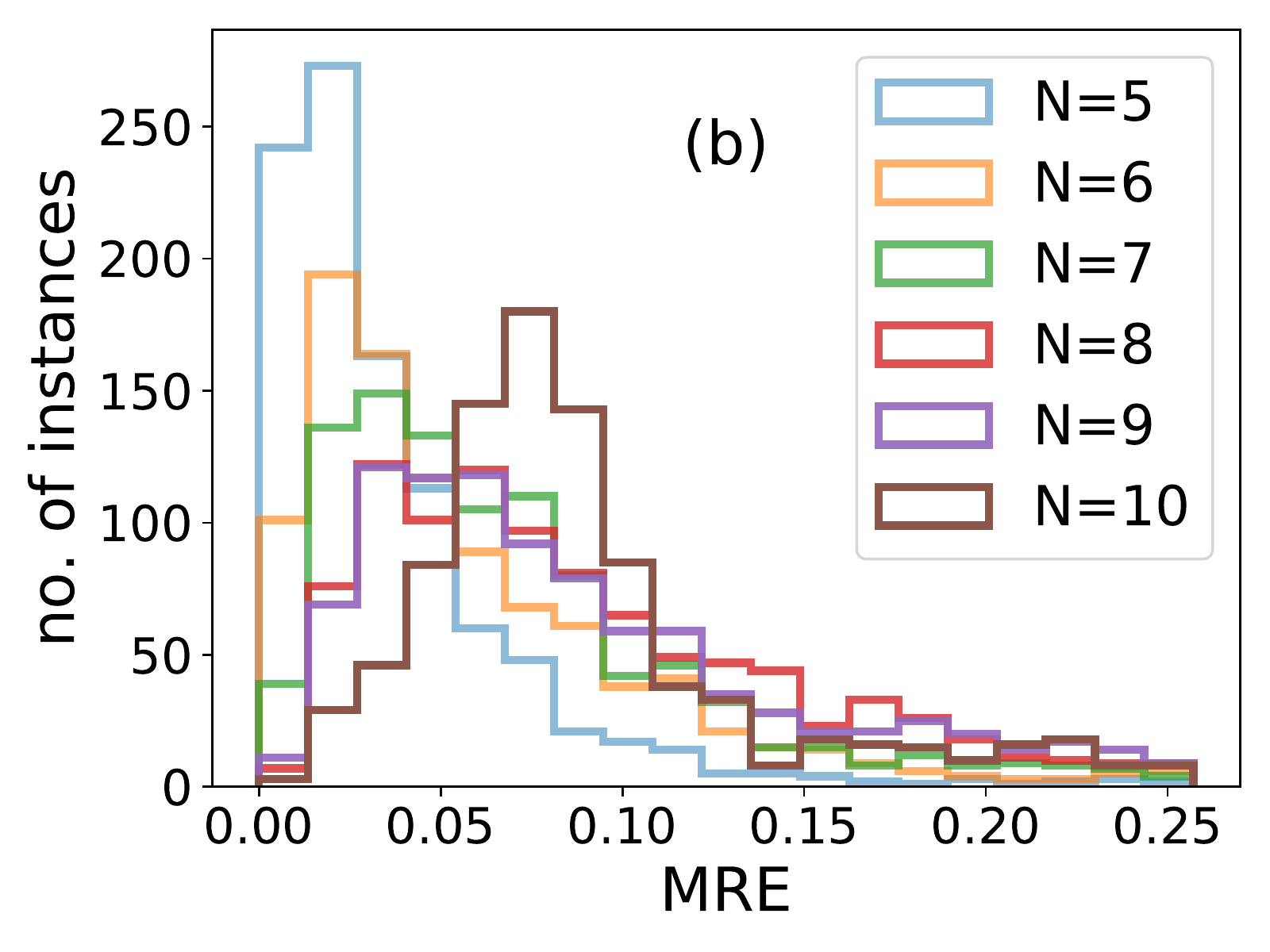}
    \includegraphics[width=0.30\textwidth]{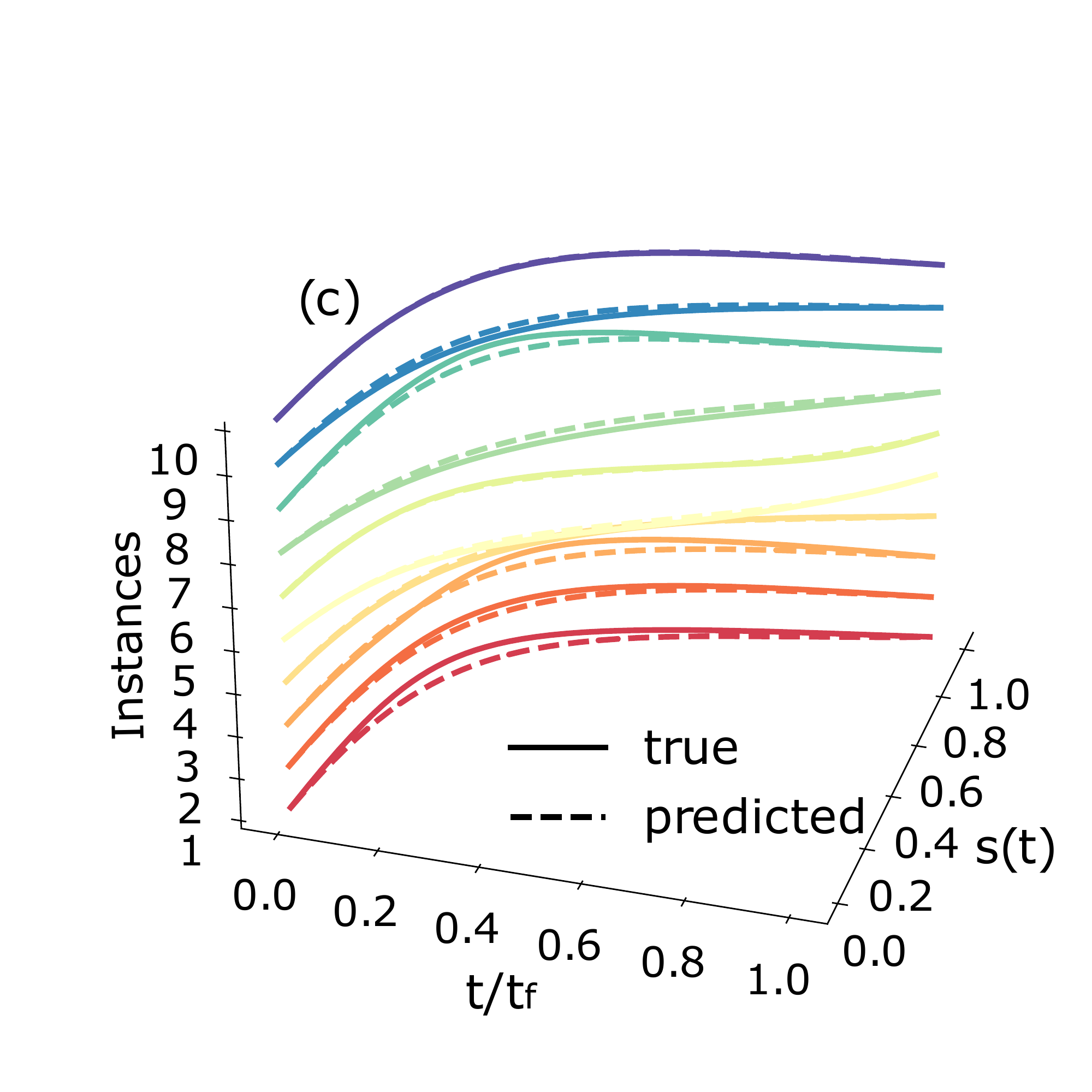}
    \caption{Training of LSTM neural network for random Ising models. (a) \removal{The mean squared error loss through the number of epochs for random Ising models with $N=5$ spins, for nearest neighbour, next-nearest neighbour model and the all-to-all connected model (see text). The loss metric is evaluated for both training data and validation data sets.}\addition{The mean relative error through the number of epochs for random Ising models with $N=5$ spins, for nearest neighbour, next-nearest neighbour model and the all-to-all connected model (see text). The metric is evaluated for both training data and validation data sets.} (b) \removal{The mean squared error loss for 1000 test instances of nearest neighbor random models with $N\in [5,10]$.}\addition{The mean relative error of prediction for 1000 test instances of nearest neighbor random models with $N\in [5,10]$.} (c) Ten test instances of annealing schedules of nearest neighbour models and the predictions made by the LSTM neural network.}
    \label{fig:lstm-training}
\end{figure*}

\section{\label{sec:problem definition}Model Hamiltonian and adiabatic conditions}
We consider random Ising models with random external magnetic fields of the form
\begin{equation}\label{eq:ising-model}
          H_\text{I} = -\sum\limits_{i} h_i\sigma_i^z+ \sum_{ ij}\frac{J_{ij}\left(\mathbb{1}-\sigma_i^z\sigma_j^z \right)}{2},
\end{equation}
 where $h_i$ represent the strength of external magnetic fields and $J_{ij}$ represent the random couplings. The second part of the Hamiltonian with random couplings is traditionally used to encode the solution to weighted Max-Cut problem~\cite{Sarjala:random-qa-ising,suzuki:qa-random-ising,crooks:qaoa, farhi:qaoa}. Max-Cut problem aims at finding a partition of a graph composed of vertices and weighted edges, such that the sum of the weights of edges connecting the two partitions is maximum. We consider longitudinal fields in addition to the traditional weighted Max-Cut Hamiltonian, in order to be more general in our treatment of random Ising models. In this manuscript, we focus on the cases of nearest neighbour interactions, next-nearest neighbour models and all-to-all connected models. The Hamiltonian couplings $h_i$ and $J_{ij}$ are chosen to be random within the discrete set with $h_i, J_{i,j} \in [-1 J_0, -0.8 J_0, ..., 0.8 J_0, 1 J_0]$, where $J_0$ defines our energy units. Times will be then measured  in units of $1/J_0$ ($\hbar=1$).
Since we consider finite-size systems, the hardest problems in this regime are the ones with smaller Hamiltonian norm. 
 % REPETITION
 %The system is initially prepared in the ground state of the Hamiltonian $H_x$ that is annealed to the given random Ising model at the final time. The time-dependent Hamiltonian which represents the process of quantum annealing is,
%
 %\begin{equation}\label{eq:qa}
 %    H(s)= A(s) H_x + B(s) H_z,
 %\end{equation}
%
%where $H_x=\sum\limits_i \sigma_i^x$ and $H_z=H_I$. $A(s)$ and $B(s)$ are time dependent annealing schedules with $s(t) \in [0, 1]$. 
%In this article, we aim at training a neural network with analytical solutions to the optimal annealing schedules for a given set of random Ising models and predict the optimal schedules for a new test random Ising model. The analytical solutions of optimal schedules are obtained from imposing local adiabatic conditions by following Roland-Cerf protocol~\cite{local-adiab:roland-cerf}, which we briefly discussed here:

The adiabatic theorem of quantum mechanics gives the global adiabaticity condition: when the system which is prepared in a ground state of the initial Hamiltonian at time $t=0$ is driven adiabatically, the final state of the system is the ground state of the final Hamiltonian with a probability
\begin{equation}
    |\langle E_0 (\tf) | \psi_0(\tf)\rangle |^2 \ge 1-\epsilon^2,
\end{equation}
where $| E_0 (\tf) \rangle$ is the exact ground state of the final Hamiltonian $\hamz$ and $|\psi_0(\tf)\rangle$ is the evolved state of the system at the final time $\tf$. This final ground state probability can be achieved provided that
 \begin{equation}\label{eq:global-adiab}
    \max\limits_{0\le t \le \tf}\left(\max\limits_{m\in \mathcal{M}}\frac{D_{0, m}}{g_{0, m}^2} \right)\le \epsilon,
 \end{equation}
where $D_{0, m}=\left\vert\left\langle m \vert \frac{dH}{dt} \vert 0\right\rangle\right\vert$ and $g_{0, m}= [E_m(t)-E_0(t)]$ is the energy gap between the ground state and the $m^{\text{th}}$ excited state. Though Eq.~\eqref{eq:global-adiab} involves in principle the entire spectrum ($\mathcal{M} \equiv [1, 2, \dots, 2^N-1]$), it is possible to focus on low-energy eigenstates only ($\mathcal{M} \equiv [1, 2, \dots, m_\text{max}]$ with $m_\text{max} < 2^N - 1$) since the system is more likely to transition to states that are closer to the ground state. In Ref.~\cite{local-adiab:roland-cerf}, the authors have introduced a more stringent, local adiabaticity condition to be verified at each time, as opposed to the global adiabatic condition of Eq.~\eqref{eq:global-adiab}. This improves the performance of adiabatic quantum computation in comparison to following the global adiabatic condition. In particular, the authors are able to achieve quadratic speedup in the adiabatic search problem.

The local adiabaticity condition for every instant of time $t$ can be stated as
\begin{equation}
    \frac{ds}{dt} \le \epsilon \min\limits_{m\in  \mathcal{M}}\left(\frac{g^2_{0,m}(s)}{\left\vert\left\langle \frac{dH}{ds}\right\rangle_{m, 0}\right\vert}\right).
\end{equation}
Integrating the above equation, we obtain the annealing time as a function of $s$:
\begin{equation}\label{eq:ts}
    t(s) = \frac{1}{\epsilon}\int_0^s \max_{m\in \mathcal{M}}\left(\frac{\left\vert\left\langle \frac{dH}{ds'}\right\rangle_{m, 0}\right\vert}{ g^2_{0,m}(s')}\right) ds'.
\end{equation}
In the numerator of the integrand, $ \frac{dH}{ds'}= H_z-H_x$, therefore we use the upper bound of the numerator of the given Ising model, $\max\limits_{m, s}\left\vert\left\langle H_z-H_x\right\rangle_{m, 0} \right\vert$ and spare it from integration. Since random Ising models can have degenerate ground states, we numerically integrate the denominator by sampling the gaps $g_{0,m}(s)$ between the ground state and the $m^\text{th}$ non-degenerate excited state, at 500 equidistant steps of $s$ between $0$ and $1$. We aim at suppressing the transitions up to $4^\text{th}$ non-degenerate excited states. We fix the value of $\epsilon=0.1$ Further, we invert the resultant function $t(s)$ to obtain the optimal annealing schedule $s(t)$ which is used to perform quantum annealing and to train the neural network. Computing a schedule requires to perform numerical integration as described in Eq.~\eqref{eq:ts}. Therefore, it is a more complicated problem that we are posing to the neural network than finding out the energy gaps as done in~\cite{lstm-sch:mohseni}. 

\addition{Once the optimal adiabatic schedule $s(t)$ has been determined, either numerically inverting Eq.~\eqref{eq:ts} or using the neural network as discussed in the next section, we use it to run a quantum annealing simulation. We build the time-dependent Hamiltonian $H(s) = [1-s(t)] H_x + s(t) H_\text{I}$. We initialize the system in the ground state of $H_x$ and use the \texttt{mesolve} integrator built in the QuTiP library~\cite{qutip} to solve the corresponding time-dependent Schr\"odinger equation, with a total annealing time given by $t(s = 1)$, see Eq.~\eqref{eq:ts}.}

%We train the LSTM neural network using a training set of  $N_\text{train}$ random Ising models in Eq.~\eqref{eq:ising-model}. We feed the Hamiltonian parameters $h_i$ and $J_{ij}$ as inputs and minimize the output error with the locally adiabatic protocol computed for the respective Ising model. 

\begin{figure*}[tb]
    \centering
    \includegraphics[width=\linewidth]{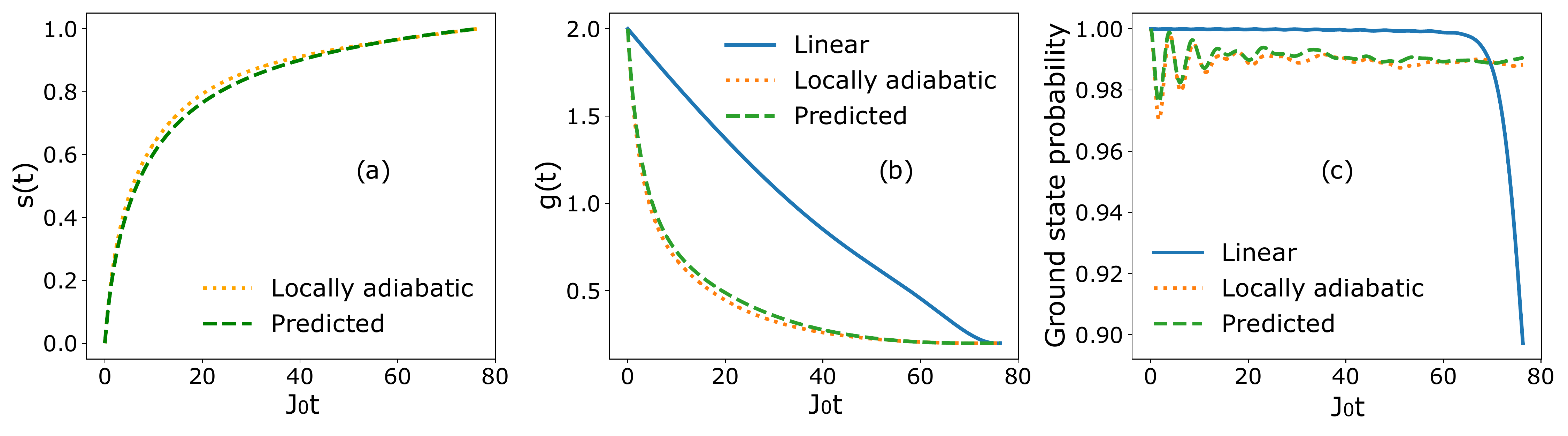}
    \caption{The results of performance of quantum annealing following the locally-adiabatic protocol of Eq.~\eqref{eq:ts} for a nearest neighbour random Ising model with $N=5$ spins. The data shown are for a test case and the results predicted by the LSTM neural network are also shown. (a)Optimal schedule obtained by imposing the local adiabaticity condition as shown in~\cite{local-adiab:roland-cerf}, and the schedule predicted by the trained LSTM network. The MSE of prediction for this schedule is $\approx 1.3 \times 10^{-5}$. (b)Evolution of minimum energy gap when evolved with linear schedule, with the optimal schedule and with the predicted schedule for the same annealing time. (c) Evolution of the instantaneous ground state probability under linear evolution of $s(t)$ and the optimal evolution for the same test instance.}
    \label{fig:qa-roland-cerf}
\end{figure*}

\begin{figure}[tb]
    \centering
    \includegraphics[width=\linewidth]{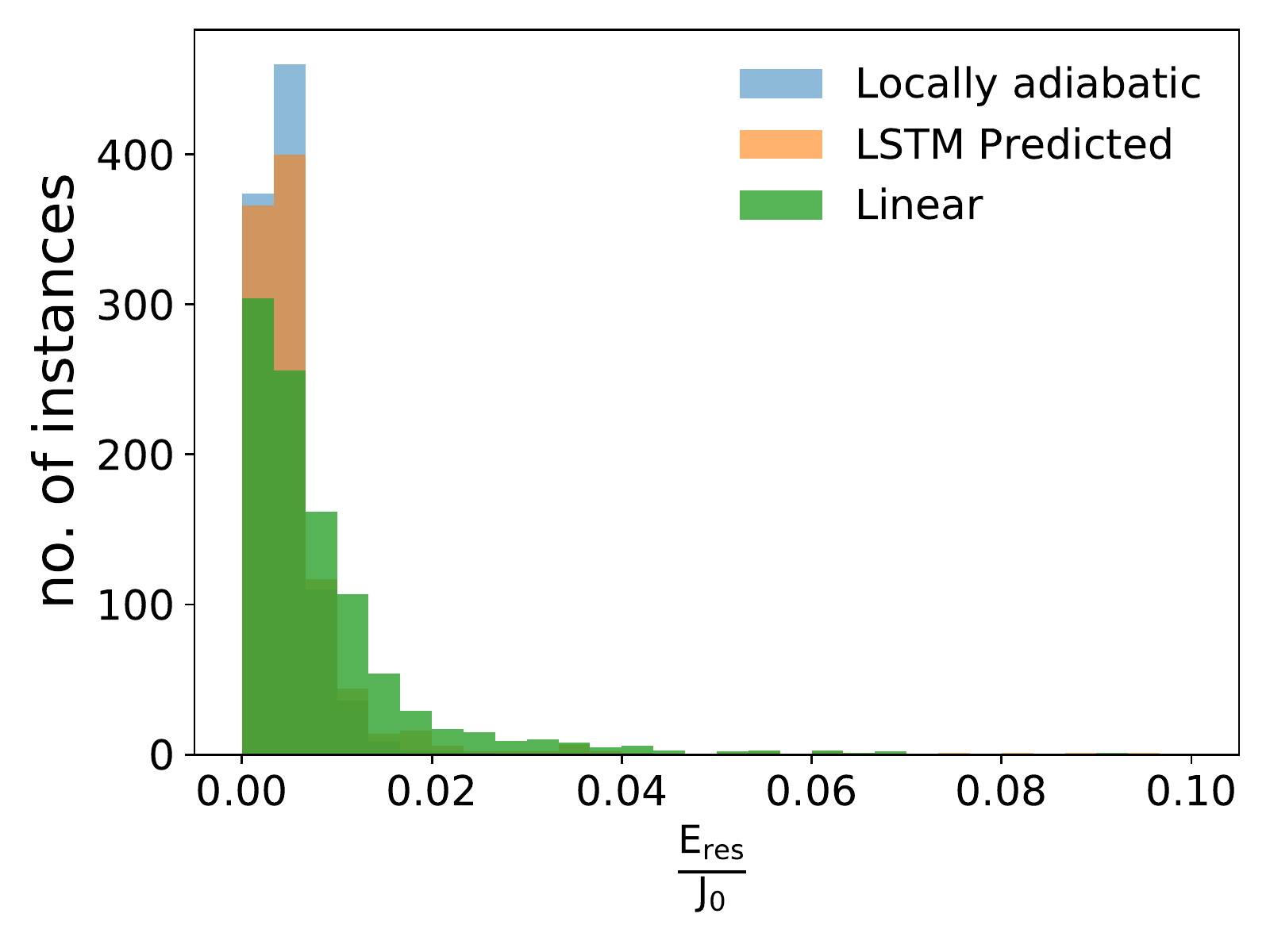}
    \caption{The distribution of the final relative residual energy with linear schedules and the optimal schedules computed using Roland and Cerf protocol and the ones predicted by LSTM neural network respectively.}
    \label{fig:app-ratios}
\end{figure}

\section{\label{sec:neural-network} Neural network training}
    Long Short-Term Memory (LSTM) neural network is a variant of Recurrent Neural Network (RNN). These types of neural networks are mainly designed to identify patterns in the sequential data. The input to these LSTM neural networks is generally a sequential data such as time series data which is passed through several layers of LSTM cells. Each LSTM cell receives information not only from the current input, but also from the previous inputs as shown in Fig.~\ref{fig:lstm}. In general, neural networks learn by minimizing the \removal{error}\addition{loss} function (such as mean squared error) between the output and the true prediction (which is known for training data set), through back propagation process. In this process, gradients of the \removal{error}\addition{loss} function are computed with respect to the weights and biases of the network. In case of recurrent neural networks, a variant algorithm called time back propagation is used~\cite{werbos:backpropagation}. However, when the sequences are long, the gradient with respect to the error may become vanishingly small, making the neural network not learn significantly through even several iterations or epochs. LSTM has special steps inside the given LSTM cell to avoid vanishing gradient problem of the classic RNN~\cite{lstm:hochreiter, lstm-sch:mohseni, koolstra:lstm}. 
    
    In this paper, our goal is to provide the information of longitudinal fields and the couplings of random Ising models as the input to the LSTM neural network and train the weights and biases of the neural network such that the output layer has the values of the optimal schedule $s(t)$ at $N_{points}$ points of time. We train the LSTM neural network with the optimal annealing schedules derived from the local adiabaticity condition given in Eq.~\eqref{eq:ts}. The neural network adjusts its weights through multiple epochs such that the mean squared error (MSE) of the output with the training data is minimum. MSE for the training data set is defined as
    \begin{equation}
    \mathrm{MSE}=\frac{1}{N_{total}}\sum\limits_{i=1}^{N_{total}} \left(y_i^{true} -y_i^{predict} \right)^2,
    \end{equation}
    where $N_{total}=N_{train}\times N_{points}$. In addition to learning, the neural network uses a small portion of training data set as a validation set of size $N_{valid}$ which is used to analyze the performance of the neural network on a data which it has never seen before. In every epoch MSE is computed for the validation data as well. For a neural network to make good predictions for new test instances, the \removal{MSE}\addition{loss function} computed over training data set and validation data set should be comparable to each other at the end of iterations. If it is not the case, it may be a signal of overfitting/underfitting, which can be addressed by adjusting the parameters of the network~\cite{mitchell_machine_1997}. \addition{Although we use MSE as the loss function for optimization for our LSTM neural network, we use the metric of mean relative error (MRE) for the further analysis of the results as MRE effectively describes the accuracy of prediction with respect to the original scale of the true schedule point which varies in range $[0, 1] $. MRE is defined as
    \begin{equation}
    \textrm{MRE}=\frac{1}{N_{points}}\sum\limits_{i=1}^{N_{points}}\frac{|y_i^{true} -y_i^{predict}| }{y^{true}_i}.
    \end{equation}
    }
\begin{figure*}[tb]
    \centering
    \includegraphics[width=\linewidth]{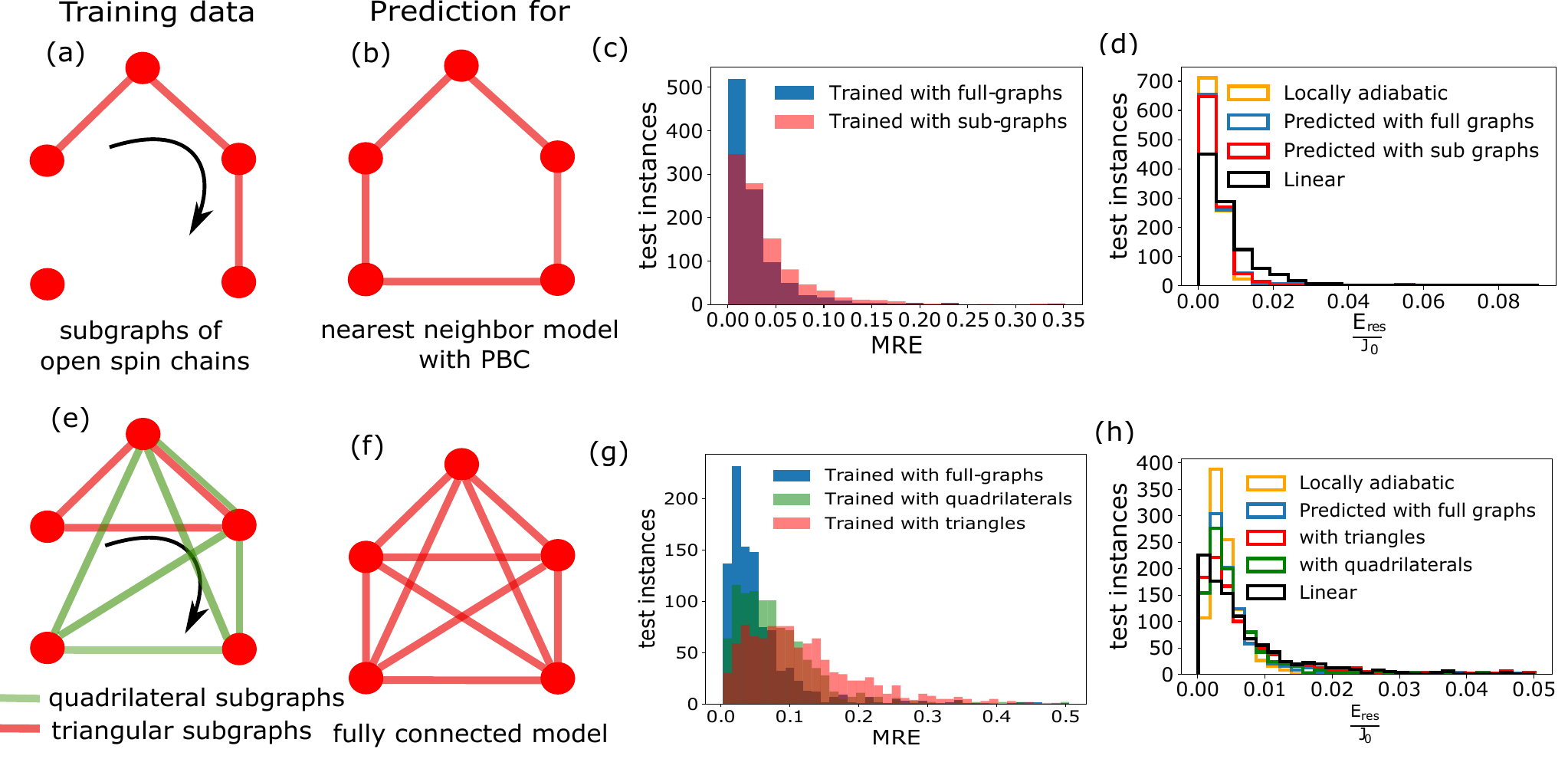}
    \caption{Optimal schedule prediction results for random Ising models with 5 spins, when trained with smaller models. (a) The neural network is trained with open spin chain with 4 spins, as shown in the graph. This trained neural network is used to make predictions of optimal schedules for the 5 spin random Ising model with periodic boundary conditions as shown in (b). In (c), we show the prediction accuracy by the neural network trained with sub-graphs in (a) and full-graphs in (b) in terms of MRE for 1000 test instances of 5 spin model, (d) the final residual energies obtained by performing quantum annealing with these predicted schedules, along with the corresponding locally adiabatic schedules and linear schedules. Similarly, in (e) we show the training data of triangular sub-graphs \addition{and quadrilateral sub-graphs} to make predictions for fully connected model in (f) and we show the distribution of MRE for 1000 instances, when predicted by neural network trained with with triangular sub-graphs\addition{, fully connected quadrilateral sub-graphs} and with fully connected models in (g). In (h) we make comparisons of residual energies obtained with different schedules specified.}
    \label{fig:extrapolation1}
\end{figure*}    
    \paragraph*{Specifications} We construct neural network with 3 LSTM layers each with 500 hidden units. Between each layer there are dropout layers with the rate 0.2, which drops $20 \%$ of the nodes randomly in each iteration and helps to overcome overfitting by reducing the complexity of the model. Each LSTM cell has tanh and sigmoid activation functions which induces non-linearity in the model~\cite{farzad:lstm-activations}. The neural network uses ``Adam" optimizer (Adaptive Moment Estimation) with the learning rate 0.0001~\cite{kingma:adam}. The number of layers, the learning rate and the number of hidden units in each layer are tuned using a validation set (0.2 \% of the training data set). We have used KERAS module to simulate and train our LSTM neural networks~\cite{keras:chollet2015}.
    
    \paragraph*{Training and predictions}
    Here we discuss our results obtained regarding the training and prediction performance of the LSTM neural network. As a first step, we use same size of models and kind of interactions for both training and predictions. For example, nearest neighbour interaction models with $\nspin$  spins are used for training to make predictions on nearest neighbour models with $\nspin$ spins.

    In Fig.~\ref{fig:lstm-training}, we show the training metrics of the LSTM neural network. Here we have considered learning random Ising models with 5 spins. The input features for this neural networks are the $h_i$ and $J_{ij}$'s of Eq.~\eqref{eq:ising-model}. The number of features $N_{features}$ depends on the connectivity of the model. We have considered three cases: 1D nearest neighbour interactions ($N_{features}=2N-1$), 1D next-nearest neighbour interactions ($N_{features}= 3N-3$), and all-to-all connected random Ising models ($N_{features}=N(N+1)/2$). Further, we use $N_{train}=\SI{50000}{}$ number of instances for training the neural network for all the cases. The output layer of the LSTM neural network is a dense layer with $N_{points}=500$ nodes which are trained to represent the optimal schedule values of $s(t)$ at 500 equidistant points of time. In this paper we mainly aim at predicting the ``shape'' of the $s(t)$ curve and not the final annealing time. It is also possible to train the network using $t(s)$ and thus learn also the optimal final annealing time, but that will be the subject of further investigation.

    In Fig.~\ref{fig:lstm-training}(a) we show the evolution of \removal{MSE}\addition{MRE} for both the training set and for a validation set of $N_{valid}=0.2\times N_{train}$ instances. It is evident that the loss metric increases with the number of input features. This result was inferred in a similar context of predicting gap evolution in adiabatic dynamics by N. Mohseni \textit{et al.}~\cite{lstm-sch:mohseni}. Here the authors showed that it is difficult to predict the properties of the fully-connected model with LSTM neural networks. In all the three cases, validation loss is close to training loss which signifies the good fit of the data by the neural network. In Fig.~\ref{fig:lstm-training}(b) we demonstrate the prediction loss for 1000 test instances in terms of \removal{MSE}\addition{MRE} for nearest neighbor models with different sizes up to 10 spins. We observe a decrease in the number of instances with the lowest \removal{MSE}\addition{MRE} with increase in system size, as expected. However even for $N=10$, the instances have \removal{MSE}\addition{MRE} within a small range. {Due to the fact that our dynamics is exact in full Hilbert space, the computational time is exponential in the number of qubits $N$. As a proof of concept, in this manuscript we study up to 10 qubit systems. As a future development, we could use tensor networks to simulate larger systems~\cite{montangero_introduction_2018}.} In Fig.~\ref{fig:lstm-training}(c), we demonstrate the optimal schedules predicted by the trained LSTM neural network and the locally adiabatic Roland and Cerf schedules for 10 new test instances of nearest neighbor random Ising models with 5 spins. These examples show that the LSTM neural network performs well and correctly predicts the annealings schedules.
%We show the distribution of average relative errors between the true and predicted schedules defined as $\frac{1}{N_{points}}\sum\limits_{i=1}^{N_{points}}\frac{y_i^{true} -y_i^{predict} }{y^{true}_i}$.

\begin{figure}[tb]
    \centering
    \includegraphics[width=\linewidth]{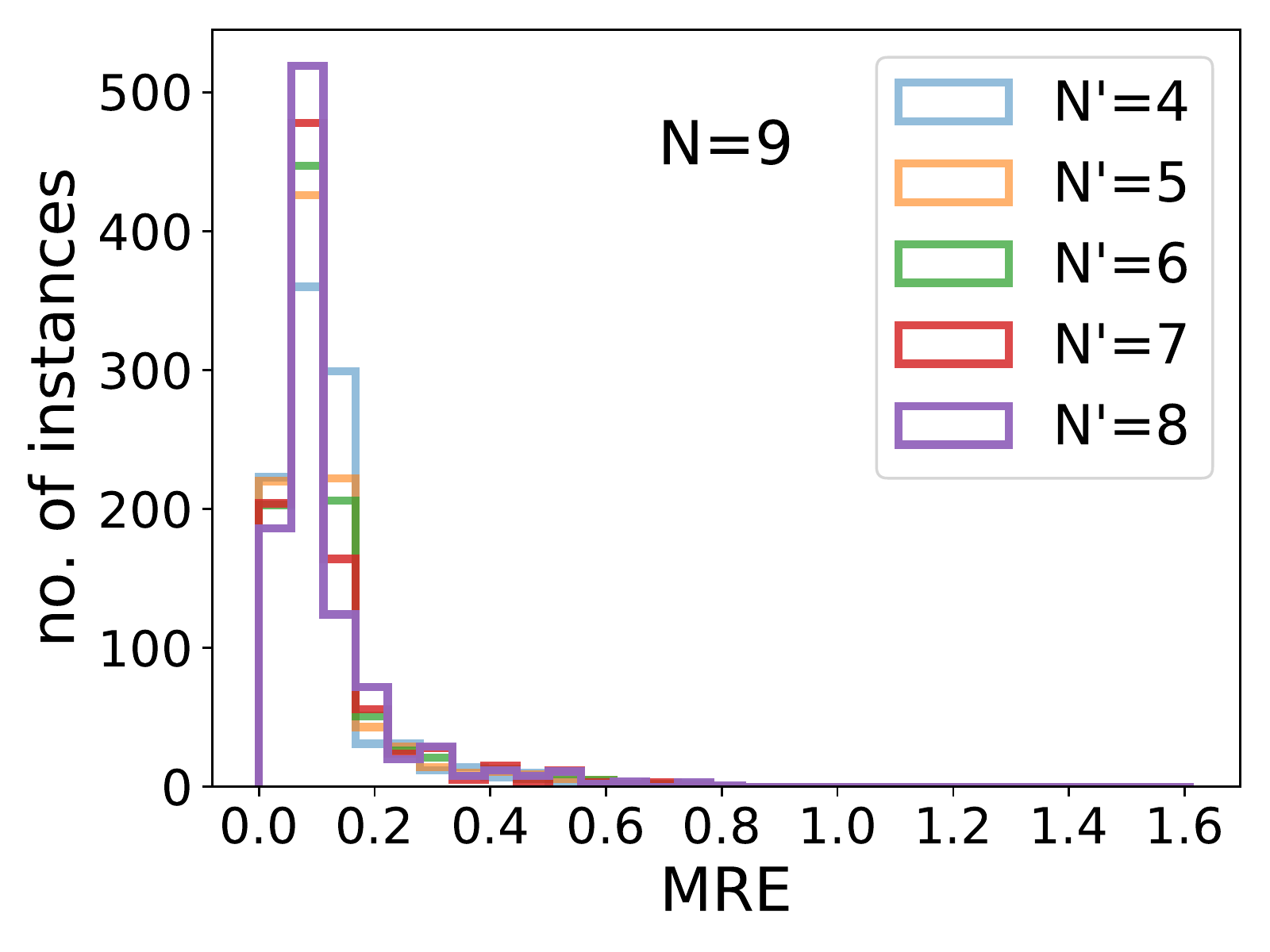}
    \caption{Prediction errors in terms of MRE for a nearest neighbour model with closed boundary condition with $N=9$. The histogram shows the distribution of errors when the neural network is trained with open chains of lengths $N'<N$.}
    \label{fig:extrapolation2}
\end{figure}
\section{\label{sec:results}Quantum annealing with LSTM Predicted annealing schedules}
Our LSTM neural network is able to predict annealing schedules that are very close to the locally-adiabatic ones in most cases. Remarkably, even when the neural network schedule shows deviations from the locally-adiabatic optimal schedule, the outcome of the network is typically a better annealing schedule compared to the linear one in the sense that  the performance of quantum annealing generally improves as a result. In this section, we show data supporting this finding.

After training the  LSTM neural network with random Ising models with  $\nspin=5$ and with nearest neighbour interactions, we predict the schedule for a new unseen test instance. In Fig.~\ref{fig:qa-roland-cerf}, we analyze the results of quantum annealing simulated with the linear schedule, Roland and Cerf schedule and the schedule predicted by the LSTM neural network. Here we consider a typical test instance of a nearest neighbour Ising model with a non degenerate ground state. In Fig.~\ref{fig:qa-roland-cerf}(a), we show the Roland and Cerf annealing schedule and the LSTM predicted schedule $s(t)$ with the final time $\tf$ predicted by Roland and Cerf protocol in Eq.~\eqref{eq:ts}, which is 76.27 (in units of $1/J_0$) for the given Ising model. Clearly, the prediction by the neural network is very accurate for this instance. In Fig.~\ref{fig:qa-roland-cerf}(b) we show the evolution of energy gap between the ground state and the first excited state following the linear schedule, optimal schedule and the predicted schedule for the given annealing time. In Fig.~\ref{fig:qa-roland-cerf}(c), we show the instantaneous ground-state probability of the system during the whole evolution, following the three protocols for the given annealing time. The optimal and the predicted schedules closely follow each other, leading to a higher ground state probability at the final time when compared to the linear schedule. 

The random Ising models can have degenerate ground states. Therefore, it is convenient to use residual energy as the measure of performance of quantum annealing. Moreover, we consider random Ising models with varying ground state energy range and hence we consider residual energy relative to the true ground state of the model, defined as
\begin{equation}
    E_{res}= \left|\frac{\langle  \psi_0(\tf)|H(s)|\psi_0(\tf)\rangle-E_0(\tf)}{E_0(\tf)}\right |.
\end{equation}
In Fig.~\ref{fig:app-ratios} we show the distribution of residual energies for the three annealing schedules for 1000 test instances of 5-spin random Ising models. Since the Ising models with smaller number of spins have considerably higher minimum energy gaps, linear schedules lead to lower residual energies as well. However, a closer look into the plots reveals that the locally adiabatic schedules and the predicted schedules have a similar distribution and have higher number of instances with lower residual energy.

%Imposing local adiabaticity conditions is one of the ways of obtaining optimal schedules which has shown success in achieving quadratic speedup in time to solution of finding the ground state. However, for a wider class of optimization problems, a more generic conditions may deliver better quantum annealing performances~\cite{kimura_rigorous_2022, Chen:optimal-adiab}. Nevertheless, the performance of LSTM neural network should be the same irrespective of the source of training data.

\section{\label{sec:extrapolation}Predicting results at large $N$, while training with $N'<N$ }
LSTM training requires a sizable training set for which one has to perform simulation of locally adiabatic schedules of different random Ising models. For systems with a large number of qubits, the computational resources required to achieve this task is daunting. As the dimension of the Hilbert space scales as $2^N$, it would be desirable if the neural network could make predictions on systems of larger dimension than the size of the systems it has been trained with. In this section we discuss the results obtained by training the neural network with the subgraphs of the given Ising model to make predictions for the full graph. We consider two cases.

\begin{figure}[tb]
    \centering
    \includegraphics[width=\linewidth]{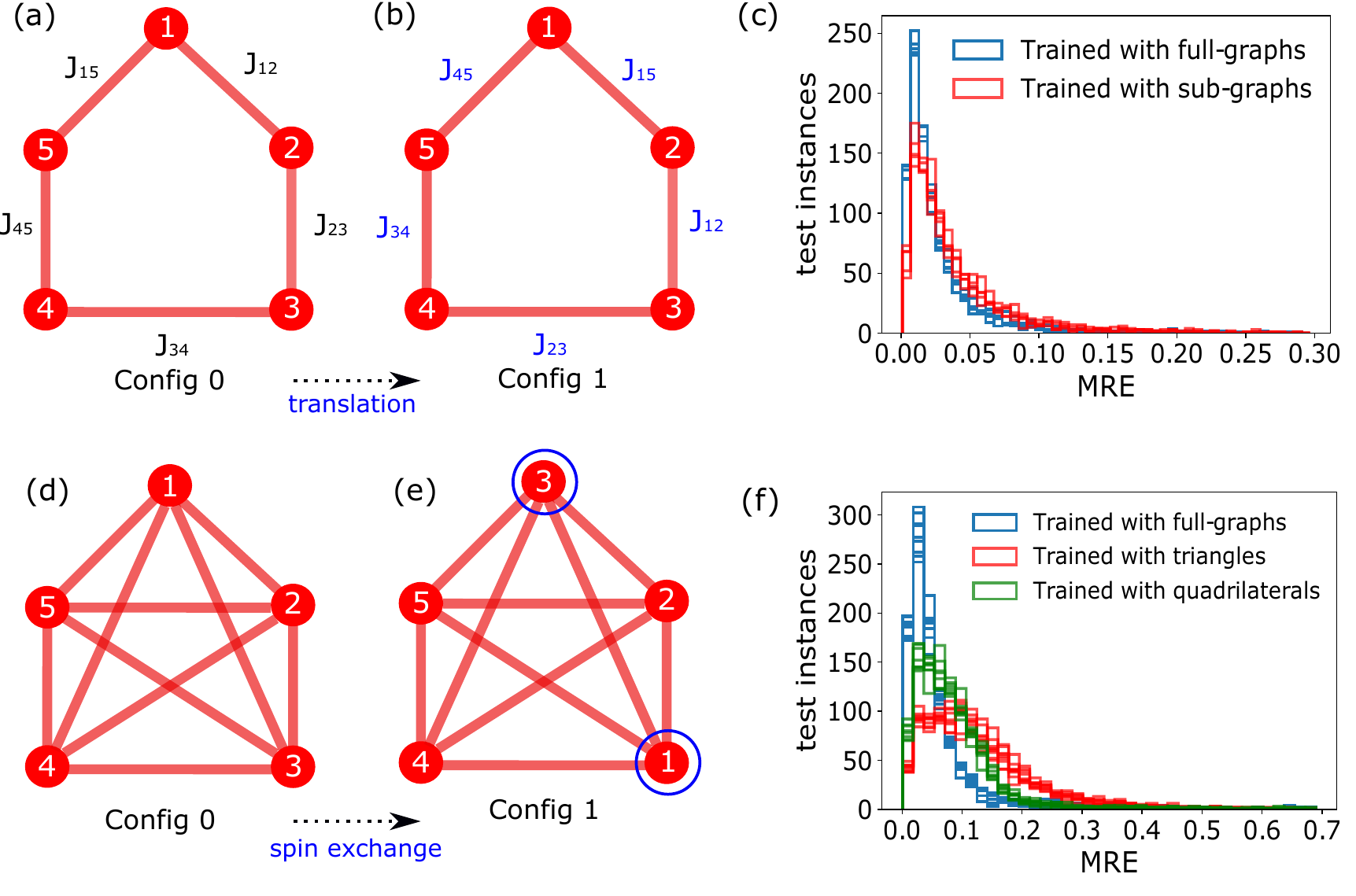}
    \caption{Prediction results of LSTM neural network for symmetric transformations of 5-spin Ising models. (a) A nearest neighbour model with periodic boundary conditions assigned as configuration 0. (b) The interactions \addition{and the local fields} are cyclically moved one step at a time to create further configuration 1. (c) the distribution of MREs for 5 configurations of translations of 1000 instances.  (d) A fully connected model referred to as configuration 0, (e) a pair of \removal{spins}\addition{spin labels (corresponding interactions and local fields)} 1 and 3 \addition{are} exchanged to create configuration 1. (f) the prediction results for 10 configurations of spin exchanges of 1000 test instances.}
    \label{fig:permutation}
\end{figure}

\begin{figure}[tb]
    \centering
    \includegraphics[width=\linewidth]{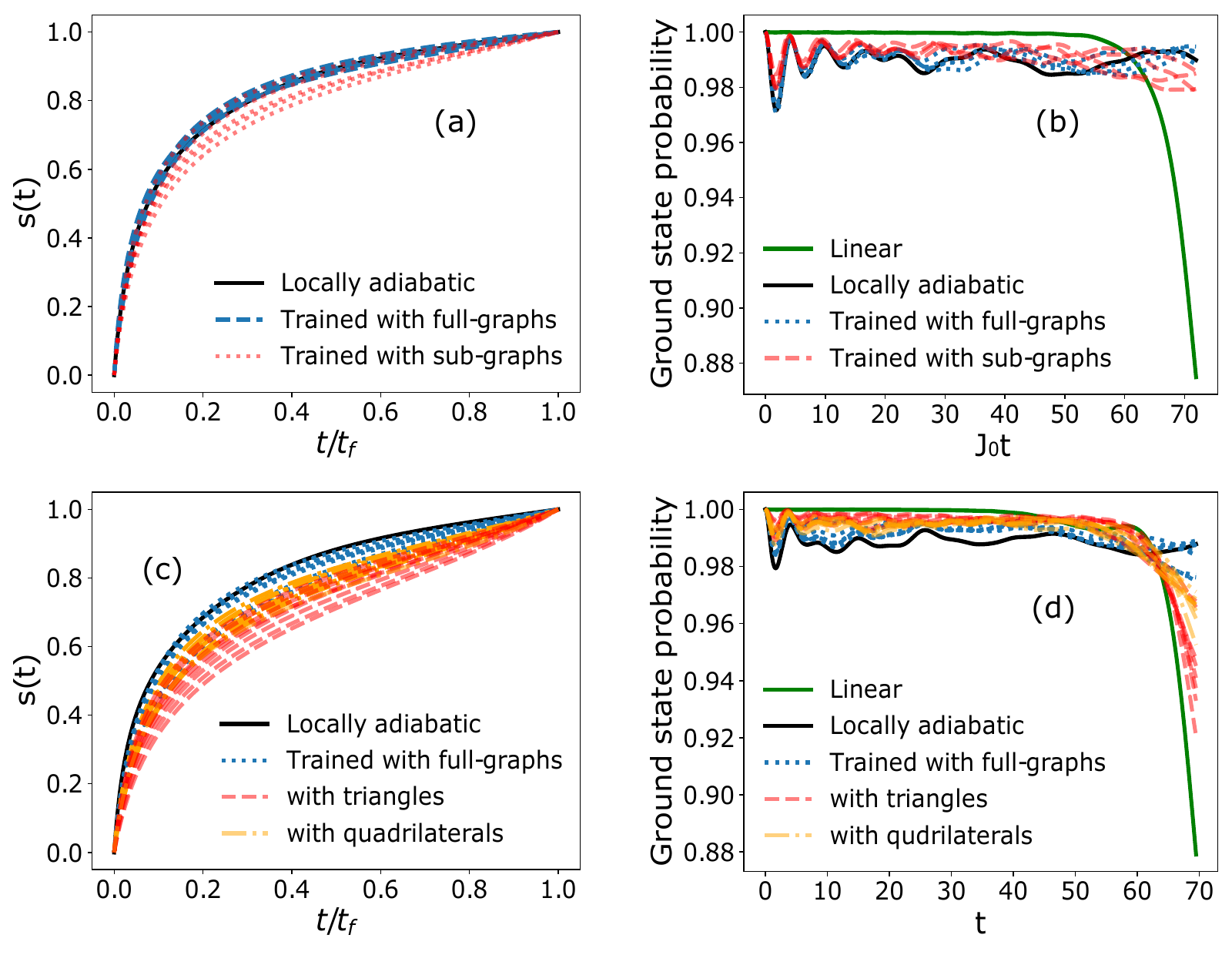}
    \caption{Depiction of annealing schedules predicted for relabelled configurations of a given model, and their performance of quantum annealing. In (a) we compare the locally adiabatic schedule and the ones predicted by the neural network, when trained with full graphs and sub graphs, for 5 configurations of a nearest neighbour model. (b) The instantaneous ground state probabilities during the adiabatic evolution following the schedules in (a). (c) and (d) show the corresponding results for a fully connected model and its 10 configurations of relabelling.  }
    \label{fig:permutation-2}
\end{figure}

\paragraph*{(a) Training over open sub-chains, predictions for larger graph (nearest-neighbors, periodic boundary conditions)}
Here we consider random Ising models with periodic boundary conditions with 5 spins. To simulate training data, we cyclically make two interactions zero at a time as shown in Fig.~\ref{fig:extrapolation1}(a). This effectively makes the training system an open chain system with $N'=4$ spins.\removal{In Fig.~\ref{fig:extrapolation1}(a), the light red lines represent the interactions which are set to zero.} We train the LSTM neural network with 5 such configurations each with \SI{40000}{} random cases. With this trained neural network, we make predictions for \removal{the} random instances of nearest neighbour model with periodic boundary conditions as shown in Fig.~\ref{fig:extrapolation1}(b). \removal{To measure the correctness of optimal schedule prediction for each instance, we choose the metric of mean relative error (MRE). For a given locally adiabatic schedule and the corresponding predicted schedule of the model, MRE is defined as}
\removal{\begin{equation}
\textrm{MRE}=\frac{1}{N_{points}}\sum\limits_{i=1}^{N_{points}}\frac{|y_i^{true} -y_i^{predict}| }{y^{true}_i}
\end{equation}}

In Fig.~\ref{fig:extrapolation1}(c), we show the distribution of \removal{mean of relative errors}\addition{MREs} for 1000 test instances of nearest neighbor models with periodic boundary conditions. The blue histogram refers to the case in which the neural network has been trained using random instances with $N = 5$ qubits, similarly to what is reported in the previous section. In this case the training data has \SI{200000} instances. The red distribution shows the prediction results when the neural network is trained with sub-graphs of different configurations. Evidently, the neural network trained with full-graphs performs slightly better with the obvious advantage of more information in the training process. It is remarkable that the neural network trained over a smaller system is able to perform comparably well to the full graph case, despite the fact that the latter encodes more information. In Fig.~\ref{fig:extrapolation1}(d), we make comparison of the performance of quantum annealing simulated the predicted schedules and the corresponding locally adiabatic and linear schedules. We observe that the schedules predicted with full-graphs and sub-graphs perform equally well and better than linear schedules.
%This is only a proof of concept, but nevertheless one can think of generalizing this idea to larger systems trained with smaller chains to further show the advantage of this approach. 
We generalize this approach to a larger system with $N=9$, while training the neural network with open spin chains with $N'=4, 5, 6, 7, 8$. We show the prediction MREs for this problem in Fig.~\ref{fig:extrapolation2}. We observe that there is no drastic increase in prediction error when the training data is of smaller size which is promising.

\paragraph*{(b) Training with triangular \removal{sub-graphs}\addition{and quadrilateral sub-graphs} and make predictions for fully connected model}
We simulate the optimal schedules for triangular sub-graphs of a fully connected  as shown in Fig.~\ref{fig:extrapolation1}(e) \addition{in red lines}, \removal{making the interactions represented by light red lines zero}. We consider 10 configurations of such triangular sub-graphs possible in the 5 spin fully connected model and we generate \SI{20000}{} random instances for each of the sub-graphs. \addition{We also consider fully connected quadrilaterals depicted in green lines of Fig.~\ref{fig:extrapolation1}(e). In this case we simulate \SI{40000}{} random instances for each of the 5 possible configurations.} We train the LSTM neural network with these data and make prediction for a fully connected model as shown in Fig.~\ref{fig:extrapolation1}(f). We show the performance of predictions in terms of MREs in Fig.~\ref{fig:extrapolation1}(g) for 1000 test instances of fully connected model when trained with triangular sub-graphs \addition{and quadrilateral sub-graphs} and compare with the prediction by the neural network trained with full-graphs of training data size 200k. \addition{LSTM trained with quadrilaterals where the training data and test data have interactions in ratio 3:5 makes better predictions than when trained with triangles where interactions in training data to test data ratio is 3:10. These results are promising because the distribution when trained with sub-graphs is peaked at small MRE despite the fact that the number of nonzero elements we feed the network is a very small fraction of the possible parameters of this fully connected graph.}\removal{These results are promising because the distribution when trained with sub-graphs is peaked at small MRE despite the fact that the number of nonzero elements we feed the network is a very small fraction of the possible parameters of this fully connected graph}. However, the tail of \removal{the graphs}\addition{histogram predicted with triangular sub-graphs} extends to higher errors when compared with the model trained with full-graphs \addition{and quadrilaterals}. As a result of worse predictions in Fig.~\ref{fig:extrapolation1} (h), we observe that the final residual energies with \addition{triangular} sub-graphs predicted schedules have similar performance as that of the linear schedules, while the schedules predicted with \addition{quadrilaterals and} full graphs lead to a better distribution of residual energies.

\section{\label{sec:permutation}Testing the invariance of predictions under translation and permutation}

In this section, we verify if the neural network is able to identify invariance of the Ising models under symmetry operations. \removal{Indeed the spectral properties of the Ising models remain invariant with the relabelling of the spins and correspondingly changing the interactions and external fields.}\addition{Relabelling the spins of a given system in an appropriate way into different configurations (correspondingly changing the interactions and external fields) represent the original system. However, for a given neural network, the relative positions of interactions and external fields in the input layer change for a relabelled model. Therefore, it is interesting to put the neural network into test to predict similar schedules for these different configurations of a system. One such meaningful operation is to rotate the labels in a cyclic manner for a system with closed boundary conditions. In case of a fully connected model, exchange of any two spin labels does not alter the spectral properties of the system.}

Firstly, we consider 5 spin models with periodic boundary conditions. We move the interaction terms cyclically to generate other configurations of this model which are symmetric in their spectral properties with respect to the original model as shown in Fig.~\ref{fig:permutation}(a). An example of the translation is shown in Fig.~\ref{fig:permutation}(b). We consider two trained neural networks to make predictions. One trained with sub-graphs of open spin chains and the one with full-graphs. Both the neural networks are trained with training data of size 200k. In Fig.~\ref{fig:permutation}(c) we show the prediction results in terms of MREs for 5 configurations by each of the neural network for 1000 test instances. For a given neural network, the predictions for all the 5 configurations have similar prediction errors. Further, the neural network trained with full-graphs of 5 spins nearest neighbour models with periodic boundary conditions (depicted in blue color) performs better than the neural network trained with sub-graphs (depicted in red color). 

As the second test, we consider a fully connected model with 5 spins and generate new configurations of the given model by exchanging \addition{labels of} a pair of spins at a time as shown in Fig.~\ref{fig:permutation}(d) and (e). \addition{We correspondingly change the interactions and local fields.} We generate 10 such configurations \addition{which essentially represent the same original system} and verify if the LSTM neural network can predict the optimal schedules for all the configurations. Again, we consider \removal{two}\addition{three} trained neural networks\removal{, one}\addition{. Two neural networks trained with triangular and quadrilateral sub-graphs respectively}\removal{trained with triangular sub-graphs} as described in the previous section and another trained with fully connected Ising models. In Fig.~\ref{fig:permutation}(f) we show the distribution of MREs of prediction for 1000 test instances. It is evident that training with full-graphs has obvious advantage over training with sub-graphs. However, MREs when trained sub-graphs peak at a considerably low error. 

We further study the performance of quantum annealing with these predicted schedules in Fig.~\ref{fig:permutation-2}. In Fig.~\ref{fig:permutation-2} (a) we consider a test instance of a nearest neighbor model with periodic boundary conditions and for a fully connected model and we show the predictions of annealing schedules for five configurations described before. Black line represents the locally adiabatic schedule for these configurations. Both the neural networks trained with full-graphs and sub-graphs make excellent prediction of  annealing schedule for all the configurations. Fig.~\ref{fig:permutation-2} (b) shows the instantaneous ground state probability with these schedules in comparison with the linear schedule. We observe that the predicted schedules drive the system with a high ground state probability until the final time, while the linear schedule drops the ground state probability of the system for the same annealing time. Similarly, in Fig.~\ref{fig:permutation-2} (c) we show a test instance of fully connected model and predictions for 10 different configurations of permutation of labels of this model. The neural network trained with full-graphs succeeds in making better predictions than the one trained with \addition{quadrilaterals followed by the one trained with}triangular sub-graphs. \removal{However, the schedules predicted with sub-graphs manages to show higher ground state probability at the final time in comparison with the linear schedule as shown in Fig.~\ref{fig:permutation-2} (d)}\addition{In Fig.~\ref{fig:permutation-2} (d) we show the performance of quantum annealing in terms of instantaneous ground state probability. Although, the schedules predicted with sub-graphs manages to show higher ground state probability during the evolution, towards the final time, they drop to a value intermediate between linear and local adiabatic schedule}. \removal{This example shows that, if the predicted schedules have a similar overall shape as that of the locally adiabatic one, they perform efficient quantum annealing despite being slightly deviated from the locally adiabatic schedule.}

%In Fig.~\ref{fig:permutation}(g) we show the prediction of annealing schedules for all the 10 configurations of a test instance. The neural network trained with full-graphs succeeds in making better predictions than the one trained with triangular sub-graphs. Moreover, for a given neural network, the predictions for all the configurations closely resemble one another signifying the ability of the neural network identify symmetric models.

The above tests show the capability of the LSTM neural networks to learn and be consistent with the symmetry properties 
of Ising models \removal{in addition to finding the optimal schedules}\addition{for nearest neighbour cases while there is room for improvement when it comes to fully connected models}.

\section{\label{sec:Conclusions}Conclusions}
In this paper we used Long Short Term Memory neural networks to learn optimal annealing schedules for random Ising models. The considered Ising models represent weighted Max-Cut problems with longitudinal fields whose interaction terms and field strength terms are chosen randomly. We followed Roland and Cerf protocol~\cite{local-adiab:roland-cerf} that consists in forcing local adiabaticity conditions to derive optimal schedules for a set of random Ising models and used them to train the neural networks. The trained models were further used to make predictions for other unseen random models.

Beginning with the same dimensional systems for both training and predictions, we were able to observe a very accurate prediction of the annealing schedules compared to that of locally adiabatic schedules. Quantum annealing performed with LSTM predicted schedules closely followed the dynamics of that of the locally adiabatic schedule whereas both these schedules outperformed the traditional linear schedule for a given annealing time. 

We showed the possibility of training the LSTM neural networks with smaller systems to make annealing schedule predictions for larger systems. The results were promising and can be hugely beneficial from the practical point of view as simulating the training data set for large systems can be computationally exhaustive. In the future, we intend to make further tests and improve our model to fit a wider class of Ising models, while training the neural network with lesser information.

As a final test, we used LSTM neural networks to make predictions for models including permutations or periodic translations of sites. % which are shuffled in their labelling with respect to each other. 
We observed that the LSTM neural network is able to predict very similar annealing schedules for these models in addition to being very close to the locally adiabatic model. Even among the cases where the schedules were slightly deviating from the locally adiabatic ones, the final ground state probability of quantum annealing with these schedules were higher than the linear schedules. This property of neural networks to predict similar schedules for symmetric models can have potential applications in employing neural networks to classify quantum systems based on their spectral properties that are kept invariant under symmetry operations. Moreover, it would be of great interest to use suitable graph neural networks which are automated to detect symmetries in the input data and thereby effectively reducing the amount of data required to train the neural networks~\cite{skolik:geometric-equivariant, wu:gnn}.

Before we conclude, we wish to point out that one of the major challenges in predicting annealing schedules is to also predict the optimal annealing time. Given that the annealing times vary within a very large range for random models, it is difficult to employ neural networks training in such scenarios. This is one of the problems that we look forward to address in the future. 

\section*{Acknowledgement}

P.R.H., G.P. and P.L. acknowledge financial support from the project PIR01 00011 “IBiSCo”, PON 2014-2020. P.L. acknowledges financial support from PNRR
MUR project CN\_00000013-ICSC, PNRR MUR project PE0000023-NQSTI as well as from the project QuantERA II Programme STAQS project that has received funding from the European Union’s Horizon 2020 research and innovation programme under Grant Agreement No 101017733. 

All the data will be available upon reasonable request.
 
%\nocite{*}

\bibliography{references}% Produces the bibliography via BibTeX.

\begin{thebibliography}{92}%
\makeatletter
\providecommand \@ifxundefined [1]{%
 \@ifx{#1\undefined}
}%
\providecommand \@ifnum [1]{%
 \ifnum #1\expandafter \@firstoftwo
 \else \expandafter \@secondoftwo
 \fi
}%
\providecommand \@ifx [1]{%
 \ifx #1\expandafter \@firstoftwo
 \else \expandafter \@secondoftwo
 \fi
}%
\providecommand \natexlab [1]{#1}%
\providecommand \enquote  [1]{``#1''}%
\providecommand \bibnamefont  [1]{#1}%
\providecommand \bibfnamefont [1]{#1}%
\providecommand \citenamefont [1]{#1}%
\providecommand \href@noop [0]{\@secondoftwo}%
\providecommand \href [0]{\begingroup \@sanitize@url \@href}%
\providecommand \@href[1]{\@@startlink{#1}\@@href}%
\providecommand \@@href[1]{\endgroup#1\@@endlink}%
\providecommand \@sanitize@url [0]{\catcode `\\12\catcode `\$12\catcode
  `\&12\catcode `\#12\catcode `\^12\catcode `\_12\catcode `\%12\relax}%
\providecommand \@@startlink[1]{}%
\providecommand \@@endlink[0]{}%
\providecommand \url  [0]{\begingroup\@sanitize@url \@url }%
\providecommand \@url [1]{\endgroup\@href {#1}{\urlprefix }}%
\providecommand \urlprefix  [0]{URL }%
\providecommand \Eprint [0]{\href }%
\providecommand \doibase [0]{https://doi.org/}%
\providecommand \selectlanguage [0]{\@gobble}%
\providecommand \bibinfo  [0]{\@secondoftwo}%
\providecommand \bibfield  [0]{\@secondoftwo}%
\providecommand \translation [1]{[#1]}%
\providecommand \BibitemOpen [0]{}%
\providecommand \bibitemStop [0]{}%
\providecommand \bibitemNoStop [0]{.\EOS\space}%
\providecommand \EOS [0]{\spacefactor3000\relax}%
\providecommand \BibitemShut  [1]{\csname bibitem#1\endcsname}%
\let\auto@bib@innerbib\@empty
%</preamble>
\bibitem [{\citenamefont {Kadowaki}\ and\ \citenamefont
  {Nishimori}(1998)}]{kadowaki-nishimori:qa}%
  \BibitemOpen
  \bibfield  {author} {\bibinfo {author} {\bibfnamefont {T.}~\bibnamefont
  {Kadowaki}}\ and\ \bibinfo {author} {\bibfnamefont {H.}~\bibnamefont
  {Nishimori}},\ }\href {https://doi.org/10.1103/PhysRevE.58.5355} {\bibfield
  {journal} {\bibinfo  {journal} {Phys. Rev. E}\ }\textbf {\bibinfo {volume}
  {58}},\ \bibinfo {pages} {5355} (\bibinfo {year} {1998})}\BibitemShut
  {NoStop}%
\bibitem [{\citenamefont {Farhi}\ \emph {et~al.}(2000)\citenamefont {Farhi},
  \citenamefont {Goldstone}, \citenamefont {Gutmann},\ and\ \citenamefont
  {Sipser}}]{Farhi:00}%
  \BibitemOpen
  \bibfield  {author} {\bibinfo {author} {\bibfnamefont {E.}~\bibnamefont
  {Farhi}}, \bibinfo {author} {\bibfnamefont {J.}~\bibnamefont {Goldstone}},
  \bibinfo {author} {\bibfnamefont {S.}~\bibnamefont {Gutmann}},\ and\ \bibinfo
  {author} {\bibfnamefont {M.}~\bibnamefont {Sipser}},\ }\href
  {http://arxiv.org/abs/quant-ph/0001106} {\bibfield  {journal} {\bibinfo
  {journal} {arXiv:quant-ph/0001106}\ } (\bibinfo {year} {2000})}\BibitemShut
  {NoStop}%
\bibitem [{\citenamefont {Brooke}\ \emph {et~al.}(1999)\citenamefont {Brooke},
  \citenamefont {Bitko}, \citenamefont {F.}, \citenamefont {Rosenbaum},\ and\
  \citenamefont {Aeppli}}]{Brooke1999}%
  \BibitemOpen
  \bibfield  {author} {\bibinfo {author} {\bibfnamefont {J.}~\bibnamefont
  {Brooke}}, \bibinfo {author} {\bibfnamefont {D.}~\bibnamefont {Bitko}},
  \bibinfo {author} {\bibfnamefont {T.}~\bibnamefont {F.}}, \bibinfo {author}
  {\bibnamefont {Rosenbaum}},\ and\ \bibinfo {author} {\bibfnamefont
  {G.}~\bibnamefont {Aeppli}},\ }\href
  {https://doi.org/10.1126/science.284.5415.779} {\bibfield  {journal}
  {\bibinfo  {journal} {Science}\ }\textbf {\bibinfo {volume} {284}},\ \bibinfo
  {pages} {779} (\bibinfo {year} {1999})}\BibitemShut {NoStop}%
\bibitem [{\citenamefont {Santoro}\ \emph {et~al.}(2002)\citenamefont
  {Santoro}, \citenamefont {Marto{\v n}{\'a}k}, \citenamefont {Tosatti},\ and\
  \citenamefont {Car}}]{santoro:qa}%
  \BibitemOpen
  \bibfield  {author} {\bibinfo {author} {\bibfnamefont {G.~E.}\ \bibnamefont
  {Santoro}}, \bibinfo {author} {\bibfnamefont {R.}~\bibnamefont {Marto{\v
  n}{\'a}k}}, \bibinfo {author} {\bibfnamefont {E.}~\bibnamefont {Tosatti}},\
  and\ \bibinfo {author} {\bibfnamefont {R.}~\bibnamefont {Car}},\ }\href
  {https://doi.org/10.1126/science.1068774} {\bibfield  {journal} {\bibinfo
  {journal} {Science}\ }\textbf {\bibinfo {volume} {295}},\ \bibinfo {pages}
  {2427} (\bibinfo {year} {2002})}\BibitemShut {NoStop}%
\bibitem [{\citenamefont {Marto\ifmmode~\check{n}\else \v{n}\fi{}\'ak}\ \emph
  {et~al.}(2004)\citenamefont {Marto\ifmmode~\check{n}\else \v{n}\fi{}\'ak},
  \citenamefont {Santoro},\ and\ \citenamefont {Tosatti}}]{PhysRevE.70.057701}%
  \BibitemOpen
  \bibfield  {author} {\bibinfo {author} {\bibfnamefont {R.}~\bibnamefont
  {Marto\ifmmode~\check{n}\else \v{n}\fi{}\'ak}}, \bibinfo {author}
  {\bibfnamefont {G.~E.}\ \bibnamefont {Santoro}},\ and\ \bibinfo {author}
  {\bibfnamefont {E.}~\bibnamefont {Tosatti}},\ }\href
  {https://doi.org/10.1103/PhysRevE.70.057701} {\bibfield  {journal} {\bibinfo
  {journal} {Phys. Rev. E}\ }\textbf {\bibinfo {volume} {70}},\ \bibinfo
  {pages} {057701} (\bibinfo {year} {2004})}\BibitemShut {NoStop}%
\bibitem [{\citenamefont {Battaglia}\ \emph {et~al.}(2005)\citenamefont
  {Battaglia}, \citenamefont {Santoro},\ and\ \citenamefont
  {Tosatti}}]{PhysRevE.71.066707}%
  \BibitemOpen
  \bibfield  {author} {\bibinfo {author} {\bibfnamefont {D.~A.}\ \bibnamefont
  {Battaglia}}, \bibinfo {author} {\bibfnamefont {G.~E.}\ \bibnamefont
  {Santoro}},\ and\ \bibinfo {author} {\bibfnamefont {E.}~\bibnamefont
  {Tosatti}},\ }\href {https://doi.org/10.1103/PhysRevE.71.066707} {\bibfield
  {journal} {\bibinfo  {journal} {Phys. Rev. E}\ }\textbf {\bibinfo {volume}
  {71}},\ \bibinfo {pages} {066707} (\bibinfo {year} {2005})}\BibitemShut
  {NoStop}%
\bibitem [{\citenamefont {Matsuda}\ \emph {et~al.}(2009)\citenamefont
  {Matsuda}, \citenamefont {Nishimori},\ and\ \citenamefont
  {Katzgraber}}]{Matsuda:2009uq}%
  \BibitemOpen
  \bibfield  {author} {\bibinfo {author} {\bibfnamefont {Y.}~\bibnamefont
  {Matsuda}}, \bibinfo {author} {\bibfnamefont {H.}~\bibnamefont {Nishimori}},\
  and\ \bibinfo {author} {\bibfnamefont {H.~G.}\ \bibnamefont {Katzgraber}},\
  }\href {http://stacks.iop.org/1367-2630/11/i=7/a=073021} {\bibfield
  {journal} {\bibinfo  {journal} {{New~J.~Phys.}}\ }\textbf {\bibinfo {volume}
  {11}},\ \bibinfo {pages} {073021} (\bibinfo {year} {2009})}\BibitemShut
  {NoStop}%
\bibitem [{\citenamefont {Perdomo-Ortiz}\ \emph {et~al.}(2012)\citenamefont
  {Perdomo-Ortiz}, \citenamefont {Dickson}, \citenamefont {Drew-Brook},
  \citenamefont {Rose},\ and\ \citenamefont
  {Aspuru-Guzik}}]{Perdomo-Ortiz2012}%
  \BibitemOpen
  \bibfield  {author} {\bibinfo {author} {\bibfnamefont {A.}~\bibnamefont
  {Perdomo-Ortiz}}, \bibinfo {author} {\bibfnamefont {N.}~\bibnamefont
  {Dickson}}, \bibinfo {author} {\bibfnamefont {M.}~\bibnamefont {Drew-Brook}},
  \bibinfo {author} {\bibfnamefont {G.}~\bibnamefont {Rose}},\ and\ \bibinfo
  {author} {\bibfnamefont {A.}~\bibnamefont {Aspuru-Guzik}},\ }\href
  {https://doi.org/10.1038/srep00571} {\bibfield  {journal} {\bibinfo
  {journal} {Scientific Reports}\ }\textbf {\bibinfo {volume} {2}},\ \bibinfo
  {pages} {571} (\bibinfo {year} {2012})}\BibitemShut {NoStop}%
\bibitem [{\citenamefont {Bian}\ \emph {et~al.}(2013)\citenamefont {Bian},
  \citenamefont {Chudak}, \citenamefont {Macready}, \citenamefont {Clark},\
  and\ \citenamefont {Gaitan}}]{Ramsey-expt}%
  \BibitemOpen
  \bibfield  {author} {\bibinfo {author} {\bibfnamefont {Z.}~\bibnamefont
  {Bian}}, \bibinfo {author} {\bibfnamefont {F.}~\bibnamefont {Chudak}},
  \bibinfo {author} {\bibfnamefont {W.~G.}\ \bibnamefont {Macready}}, \bibinfo
  {author} {\bibfnamefont {L.}~\bibnamefont {Clark}},\ and\ \bibinfo {author}
  {\bibfnamefont {F.}~\bibnamefont {Gaitan}},\ }\href
  {http://link.aps.org/doi/10.1103/PhysRevLett.111.130505} {\bibfield
  {journal} {\bibinfo  {journal} {Physical Review Letters}\ }\textbf {\bibinfo
  {volume} {111}},\ \bibinfo {pages} {130505} (\bibinfo {year}
  {2013})}\BibitemShut {NoStop}%
\bibitem [{\citenamefont {R{\o}nnow}\ \emph {et~al.}(2014)\citenamefont
  {R{\o}nnow}, \citenamefont {Wang}, \citenamefont {Job}, \citenamefont
  {Boixo}, \citenamefont {Isakov}, \citenamefont {Wecker}, \citenamefont
  {Martinis}, \citenamefont {Lidar},\ and\ \citenamefont
  {Troyer}}]{ronnow:speedup}%
  \BibitemOpen
  \bibfield  {author} {\bibinfo {author} {\bibfnamefont {T.~F.}\ \bibnamefont
  {R{\o}nnow}}, \bibinfo {author} {\bibfnamefont {Z.}~\bibnamefont {Wang}},
  \bibinfo {author} {\bibfnamefont {J.}~\bibnamefont {Job}}, \bibinfo {author}
  {\bibfnamefont {S.}~\bibnamefont {Boixo}}, \bibinfo {author} {\bibfnamefont
  {S.~V.}\ \bibnamefont {Isakov}}, \bibinfo {author} {\bibfnamefont
  {D.}~\bibnamefont {Wecker}}, \bibinfo {author} {\bibfnamefont {J.~M.}\
  \bibnamefont {Martinis}}, \bibinfo {author} {\bibfnamefont {D.~A.}\
  \bibnamefont {Lidar}},\ and\ \bibinfo {author} {\bibfnamefont
  {M.}~\bibnamefont {Troyer}},\ }\href
  {https://doi.org/10.1126/science.1252319} {\bibfield  {journal} {\bibinfo
  {journal} {Science}\ }\textbf {\bibinfo {volume} {345}},\ \bibinfo {pages}
  {420} (\bibinfo {year} {2014})}\BibitemShut {NoStop}%
\bibitem [{\citenamefont {Rieffel}\ \emph {et~al.}(2015)\citenamefont
  {Rieffel}, \citenamefont {Venturelli}, \citenamefont {O'Gorman},
  \citenamefont {Do}, \citenamefont {Prystay},\ and\ \citenamefont
  {Smelyanskiy}}]{Rieffel:2015aa}%
  \BibitemOpen
  \bibfield  {author} {\bibinfo {author} {\bibfnamefont {E.~G.}\ \bibnamefont
  {Rieffel}}, \bibinfo {author} {\bibfnamefont {D.}~\bibnamefont {Venturelli}},
  \bibinfo {author} {\bibfnamefont {B.}~\bibnamefont {O'Gorman}}, \bibinfo
  {author} {\bibfnamefont {M.~B.}\ \bibnamefont {Do}}, \bibinfo {author}
  {\bibfnamefont {E.~M.}\ \bibnamefont {Prystay}},\ and\ \bibinfo {author}
  {\bibfnamefont {V.~N.}\ \bibnamefont {Smelyanskiy}},\ }\href
  {https://doi.org/10.1007/s11128-014-0892-x} {\bibfield  {journal} {\bibinfo
  {journal} {Quantum Information Processing}\ }\textbf {\bibinfo {volume}
  {14}},\ \bibinfo {pages} {1} (\bibinfo {year} {2015})}\BibitemShut {NoStop}%
\bibitem [{\citenamefont {Azinovi{\'c}}\ \emph {et~al.}(2017)\citenamefont
  {Azinovi{\'c}}, \citenamefont {Herr}, \citenamefont {Heim}, \citenamefont
  {Brown},\ and\ \citenamefont {Troyer}}]{Azinovic:2016uq}%
  \BibitemOpen
  \bibfield  {author} {\bibinfo {author} {\bibfnamefont {M.}~\bibnamefont
  {Azinovi{\'c}}}, \bibinfo {author} {\bibfnamefont {D.}~\bibnamefont {Herr}},
  \bibinfo {author} {\bibfnamefont {B.}~\bibnamefont {Heim}}, \bibinfo {author}
  {\bibfnamefont {E.}~\bibnamefont {Brown}},\ and\ \bibinfo {author}
  {\bibfnamefont {M.}~\bibnamefont {Troyer}},\ }\href
  {https://doi.org/10.21468/SciPostPhys.2.2.013} {\bibfield  {journal}
  {\bibinfo  {journal} {SciPost Physics}\ }\textbf {\bibinfo {volume} {2}},\
  \bibinfo {pages} {013} (\bibinfo {year} {2017})}\BibitemShut {NoStop}%
\bibitem [{\citenamefont {Mott}\ \emph {et~al.}(2017)\citenamefont {Mott},
  \citenamefont {Job}, \citenamefont {Vlimant}, \citenamefont {Lidar},\ and\
  \citenamefont {Spiropulu}}]{Mott:2017aa}%
  \BibitemOpen
  \bibfield  {author} {\bibinfo {author} {\bibfnamefont {A.}~\bibnamefont
  {Mott}}, \bibinfo {author} {\bibfnamefont {J.}~\bibnamefont {Job}}, \bibinfo
  {author} {\bibfnamefont {J.-R.}\ \bibnamefont {Vlimant}}, \bibinfo {author}
  {\bibfnamefont {D.}~\bibnamefont {Lidar}},\ and\ \bibinfo {author}
  {\bibfnamefont {M.}~\bibnamefont {Spiropulu}},\ }\href
  {http://dx.doi.org/10.1038/nature24047} {\bibfield  {journal} {\bibinfo
  {journal} {Nature}\ }\textbf {\bibinfo {volume} {550}},\ \bibinfo {pages}
  {375 EP } (\bibinfo {year} {2017})}\BibitemShut {NoStop}%
\bibitem [{\citenamefont {Li}\ \emph {et~al.}(2018)\citenamefont {Li},
  \citenamefont {Di~Felice}, \citenamefont {Rohs},\ and\ \citenamefont
  {Lidar}}]{Li:comp-bio-2017}%
  \BibitemOpen
  \bibfield  {author} {\bibinfo {author} {\bibfnamefont {R.~Y.}\ \bibnamefont
  {Li}}, \bibinfo {author} {\bibfnamefont {R.}~\bibnamefont {Di~Felice}},
  \bibinfo {author} {\bibfnamefont {R.}~\bibnamefont {Rohs}},\ and\ \bibinfo
  {author} {\bibfnamefont {D.~A.}\ \bibnamefont {Lidar}},\ }\href
  {https://doi.org/10.1038/s41534-018-0060-8} {\bibfield  {journal} {\bibinfo
  {journal} {npj Quantum Information}\ }\textbf {\bibinfo {volume} {4}},\
  \bibinfo {pages} {14} (\bibinfo {year} {2018})}\BibitemShut {NoStop}%
\bibitem [{\citenamefont {Mandr{\`a}}\ and\ \citenamefont
  {Katzgraber}(2018)}]{Mandra:2017ab}%
  \BibitemOpen
  \bibfield  {author} {\bibinfo {author} {\bibfnamefont {S.}~\bibnamefont
  {Mandr{\`a}}}\ and\ \bibinfo {author} {\bibfnamefont {H.~G.}\ \bibnamefont
  {Katzgraber}},\ }\href {https://doi.org/10.1088/2058-9565/aac8b2} {\bibfield
  {journal} {\bibinfo  {journal} {Quantum Sci. Technol.}\ }\textbf {\bibinfo
  {volume} {3}},\ \bibinfo {pages} {04LT01} (\bibinfo {year}
  {2018})}\BibitemShut {NoStop}%
\bibitem [{\citenamefont {Jiang}\ \emph {et~al.}(2018)\citenamefont {Jiang},
  \citenamefont {Britt}, \citenamefont {McCaskey}, \citenamefont {Humble},\
  and\ \citenamefont {Kais}}]{Jiang2018}%
  \BibitemOpen
  \bibfield  {author} {\bibinfo {author} {\bibfnamefont {S.}~\bibnamefont
  {Jiang}}, \bibinfo {author} {\bibfnamefont {K.~A.}\ \bibnamefont {Britt}},
  \bibinfo {author} {\bibfnamefont {A.~J.}\ \bibnamefont {McCaskey}}, \bibinfo
  {author} {\bibfnamefont {T.~S.}\ \bibnamefont {Humble}},\ and\ \bibinfo
  {author} {\bibfnamefont {S.}~\bibnamefont {Kais}},\ }\href
  {https://doi.org/10.1038/s41598-018-36058-z} {\bibfield  {journal} {\bibinfo
  {journal} {Scientific Reports}\ }\textbf {\bibinfo {volume} {8}},\ \bibinfo
  {pages} {17667} (\bibinfo {year} {2018})}\BibitemShut {NoStop}%
\bibitem [{\citenamefont {Venturelli}\ and\ \citenamefont
  {Kondratyev}(2019{\natexlab{a}})}]{Venturelli2019}%
  \BibitemOpen
  \bibfield  {author} {\bibinfo {author} {\bibfnamefont {D.}~\bibnamefont
  {Venturelli}}\ and\ \bibinfo {author} {\bibfnamefont {A.}~\bibnamefont
  {Kondratyev}},\ }\href {https://doi.org/10.1007/s42484-019-00001-w}
  {\bibfield  {journal} {\bibinfo  {journal} {Quantum Machine Intelligence}\
  }\textbf {\bibinfo {volume} {1}},\ \bibinfo {pages} {17} (\bibinfo {year}
  {2019}{\natexlab{a}})}\BibitemShut {NoStop}%
\bibitem [{\citenamefont {Smelyanskiy}\ \emph {et~al.}(2020)\citenamefont
  {Smelyanskiy}, \citenamefont {Kechedzhi}, \citenamefont {Boixo},
  \citenamefont {Isakov}, \citenamefont {Neven},\ and\ \citenamefont
  {Altshuler}}]{Smelyanskiy:2018aa}%
  \BibitemOpen
  \bibfield  {author} {\bibinfo {author} {\bibfnamefont {V.~N.}\ \bibnamefont
  {Smelyanskiy}}, \bibinfo {author} {\bibfnamefont {K.}~\bibnamefont
  {Kechedzhi}}, \bibinfo {author} {\bibfnamefont {S.}~\bibnamefont {Boixo}},
  \bibinfo {author} {\bibfnamefont {S.~V.}\ \bibnamefont {Isakov}}, \bibinfo
  {author} {\bibfnamefont {H.}~\bibnamefont {Neven}},\ and\ \bibinfo {author}
  {\bibfnamefont {B.}~\bibnamefont {Altshuler}},\ }\href
  {https://doi.org/10.1103/PhysRevX.10.011017} {\bibfield  {journal} {\bibinfo
  {journal} {Physical Review X}\ }\textbf {\bibinfo {volume} {10}},\ \bibinfo
  {pages} {011017} (\bibinfo {year} {2020})}\BibitemShut {NoStop}%
\bibitem [{\citenamefont {Zlokapa}\ \emph {et~al.}(2021)\citenamefont
  {Zlokapa}, \citenamefont {Anand}, \citenamefont {Vlimant}, \citenamefont
  {Duarte}, \citenamefont {Job}, \citenamefont {Lidar},\ and\ \citenamefont
  {Spiropulu}}]{Zlokapa:2019ab}%
  \BibitemOpen
  \bibfield  {author} {\bibinfo {author} {\bibfnamefont {A.}~\bibnamefont
  {Zlokapa}}, \bibinfo {author} {\bibfnamefont {A.}~\bibnamefont {Anand}},
  \bibinfo {author} {\bibfnamefont {J.-R.}\ \bibnamefont {Vlimant}}, \bibinfo
  {author} {\bibfnamefont {J.~M.}\ \bibnamefont {Duarte}}, \bibinfo {author}
  {\bibfnamefont {J.}~\bibnamefont {Job}}, \bibinfo {author} {\bibfnamefont
  {D.}~\bibnamefont {Lidar}},\ and\ \bibinfo {author} {\bibfnamefont
  {M.}~\bibnamefont {Spiropulu}},\ }\href
  {https://doi.org/10.1007/s42484-021-00054-w} {\bibfield  {journal} {\bibinfo
  {journal} {Quantum Machine Intelligence}\ }\textbf {\bibinfo {volume} {3}},\
  \bibinfo {pages} {27} (\bibinfo {year} {2021})}\BibitemShut {NoStop}%
\bibitem [{\citenamefont {Tanaka}\ \emph {et~al.}(2017)\citenamefont {Tanaka},
  \citenamefont {Tamura},\ and\ \citenamefont {Chakrabarti}}]{Tanaka:book}%
  \BibitemOpen
  \bibfield  {author} {\bibinfo {author} {\bibfnamefont {S.}~\bibnamefont
  {Tanaka}}, \bibinfo {author} {\bibfnamefont {R.}~\bibnamefont {Tamura}},\
  and\ \bibinfo {author} {\bibfnamefont {B.~K.}\ \bibnamefont {Chakrabarti}},\
  }\href {https://dl.acm.org/doi/10.5555/3159044} {\emph {\bibinfo {title}
  {Quantum Spin Glasses, Annealing and Computation}}},\ \bibinfo {edition}
  {1st}\ ed.\ (\bibinfo  {publisher} {Cambridge University Press},\ \bibinfo
  {address} {{Cambride, UK}},\ \bibinfo {year} {2017})\BibitemShut {NoStop}%
\bibitem [{\citenamefont {Albash}\ and\ \citenamefont
  {Lidar}(2018)}]{albash:aqc-review}%
  \BibitemOpen
  \bibfield  {author} {\bibinfo {author} {\bibfnamefont {T.}~\bibnamefont
  {Albash}}\ and\ \bibinfo {author} {\bibfnamefont {D.~A.}\ \bibnamefont
  {Lidar}},\ }\href {https://doi.org/10.1103/RevModPhys.90.015002} {\bibfield
  {journal} {\bibinfo  {journal} {Rev. Mod. Phys.}\ }\textbf {\bibinfo {volume}
  {90}},\ \bibinfo {pages} {015002} (\bibinfo {year} {2018})}\BibitemShut
  {NoStop}%
\bibitem [{\citenamefont {Hauke}\ \emph {et~al.}(2020)\citenamefont {Hauke},
  \citenamefont {Katzgraber}, \citenamefont {Lechner}, \citenamefont
  {Nishimori},\ and\ \citenamefont {Oliver}}]{Hauke:2019aa}%
  \BibitemOpen
  \bibfield  {author} {\bibinfo {author} {\bibfnamefont {P.}~\bibnamefont
  {Hauke}}, \bibinfo {author} {\bibfnamefont {H.~G.}\ \bibnamefont
  {Katzgraber}}, \bibinfo {author} {\bibfnamefont {W.}~\bibnamefont {Lechner}},
  \bibinfo {author} {\bibfnamefont {H.}~\bibnamefont {Nishimori}},\ and\
  \bibinfo {author} {\bibfnamefont {W.~D.}\ \bibnamefont {Oliver}},\ }\href
  {https://iopscience.iop.org/article/10.1088/1361-6633/ab85b8} {\bibfield
  {journal} {\bibinfo  {journal} {Reports on Progress in Physics}\ } (\bibinfo
  {year} {2020})}\BibitemShut {NoStop}%
\bibitem [{\citenamefont {Born}\ and\ \citenamefont {Fock}(1928)}]{Born1928}%
  \BibitemOpen
  \bibfield  {author} {\bibinfo {author} {\bibfnamefont {M.}~\bibnamefont
  {Born}}\ and\ \bibinfo {author} {\bibfnamefont {V.}~\bibnamefont {Fock}},\
  }\href {https://doi.org/10.1007/BF01343193} {\bibfield  {journal} {\bibinfo
  {journal} {Zeitschrift f{\"u}r Physik}\ }\textbf {\bibinfo {volume} {51}},\
  \bibinfo {pages} {165} (\bibinfo {year} {1928})}\BibitemShut {NoStop}%
\bibitem [{\citenamefont {Sarandy}\ \emph {et~al.}(2004)\citenamefont
  {Sarandy}, \citenamefont {Wu},\ and\ \citenamefont
  {Lidar}}]{lidar:adiabatic-theorem}%
  \BibitemOpen
  \bibfield  {author} {\bibinfo {author} {\bibfnamefont {M.~S.}\ \bibnamefont
  {Sarandy}}, \bibinfo {author} {\bibfnamefont {L.-A.}\ \bibnamefont {Wu}},\
  and\ \bibinfo {author} {\bibfnamefont {D.~A.}\ \bibnamefont {Lidar}},\ }\href
  {https://doi.org/10.1007/s11128-004-7712-7} {\bibfield  {journal} {\bibinfo
  {journal} {Quantum Information Processing}\ }\textbf {\bibinfo {volume}
  {3}},\ \bibinfo {pages} {331} (\bibinfo {year} {2004})}\BibitemShut {NoStop}%
\bibitem [{\citenamefont {Kato}(1950)}]{kato:adiabatic-theorem}%
  \BibitemOpen
  \bibfield  {author} {\bibinfo {author} {\bibfnamefont {T.}~\bibnamefont
  {Kato}},\ }\href {https://doi.org/10.1143/JPSJ.5.435} {\bibfield  {journal}
  {\bibinfo  {journal} {Journal of the Physical Society of Japan}\ }\textbf
  {\bibinfo {volume} {5}},\ \bibinfo {pages} {435} (\bibinfo {year} {1950})},\
  \Eprint {https://arxiv.org/abs/https://doi.org/10.1143/JPSJ.5.435}
  {https://doi.org/10.1143/JPSJ.5.435} \BibitemShut {NoStop}%
\bibitem [{\citenamefont {Jansen}\ \emph {et~al.}(2007)\citenamefont {Jansen},
  \citenamefont {Ruskai},\ and\ \citenamefont
  {Seiler}}]{jansen:adiabatic-theorem}%
  \BibitemOpen
  \bibfield  {author} {\bibinfo {author} {\bibfnamefont {S.}~\bibnamefont
  {Jansen}}, \bibinfo {author} {\bibfnamefont {M.-B.}\ \bibnamefont {Ruskai}},\
  and\ \bibinfo {author} {\bibfnamefont {R.}~\bibnamefont {Seiler}},\ }\href
  {https://doi.org/10.1063/1.2798382} {\bibfield  {journal} {\bibinfo
  {journal} {Journal of Mathematical Physics}\ }\textbf {\bibinfo {volume}
  {48}},\ \bibinfo {pages} {102111} (\bibinfo {year} {2007})}\BibitemShut
  {NoStop}%
\bibitem [{\citenamefont {J\"org}\ \emph {et~al.}(2010)\citenamefont {J\"org},
  \citenamefont {Krzakala}, \citenamefont {Kurchan}, \citenamefont {Maggs},\
  and\ \citenamefont {Pujos}}]{jorg:energy-gaps}%
  \BibitemOpen
  \bibfield  {author} {\bibinfo {author} {\bibfnamefont {T.}~\bibnamefont
  {J\"org}}, \bibinfo {author} {\bibfnamefont {F.}~\bibnamefont {Krzakala}},
  \bibinfo {author} {\bibfnamefont {J.}~\bibnamefont {Kurchan}}, \bibinfo
  {author} {\bibfnamefont {A.~C.}\ \bibnamefont {Maggs}},\ and\ \bibinfo
  {author} {\bibfnamefont {J.}~\bibnamefont {Pujos}},\ }\href
  {https://doi.org/10.1209/0295-5075/89/40004} {\bibfield  {journal} {\bibinfo
  {journal} {{EPL} (Europhysics Letters)}\ }\textbf {\bibinfo {volume} {89}},\
  \bibinfo {pages} {40004} (\bibinfo {year} {2010})}\BibitemShut {NoStop}%
\bibitem [{\citenamefont {Lucas}(2014)}]{lucas:np-complete}%
  \BibitemOpen
  \bibfield  {author} {\bibinfo {author} {\bibfnamefont {A.}~\bibnamefont
  {Lucas}},\ }\href {https://doi.org/10.3389/fphy.2014.00005} {\bibfield
  {journal} {\bibinfo  {journal} {Frontiers in Physics}\ }\textbf {\bibinfo
  {volume} {2}},\ \bibinfo {pages} {5} (\bibinfo {year} {2014})}\BibitemShut
  {NoStop}%
\bibitem [{\citenamefont {Preskill}(2018)}]{Preskill2018}%
  \BibitemOpen
  \bibfield  {author} {\bibinfo {author} {\bibfnamefont {J.}~\bibnamefont
  {Preskill}},\ }\href {https://doi.org/10.22331/q-2018-08-06-79} {\bibfield
  {journal} {\bibinfo  {journal} {Quantum}\ }\textbf {\bibinfo {volume} {2}},\
  \bibinfo {pages} {79} (\bibinfo {year} {2018})}\BibitemShut {NoStop}%
\bibitem [{\citenamefont {Albash}\ and\ \citenamefont
  {Lidar}(2015)}]{albash:decoherence}%
  \BibitemOpen
  \bibfield  {author} {\bibinfo {author} {\bibfnamefont {T.}~\bibnamefont
  {Albash}}\ and\ \bibinfo {author} {\bibfnamefont {D.~A.}\ \bibnamefont
  {Lidar}},\ }\href {http://link.aps.org/doi/10.1103/PhysRevA.91.062320}
  {\bibfield  {journal} {\bibinfo  {journal} {Phys. Rev. A}\ }\textbf {\bibinfo
  {volume} {91}},\ \bibinfo {pages} {062320} (\bibinfo {year}
  {2015})}\BibitemShut {NoStop}%
\bibitem [{\citenamefont {Amin}(2015)}]{amin:speedup}%
  \BibitemOpen
  \bibfield  {author} {\bibinfo {author} {\bibfnamefont {M.~H.}\ \bibnamefont
  {Amin}},\ }\href {https://doi.org/10.1103/PhysRevA.92.052323} {\bibfield
  {journal} {\bibinfo  {journal} {Phys. Rev. A}\ }\textbf {\bibinfo {volume}
  {92}},\ \bibinfo {pages} {052323} (\bibinfo {year} {2015})}\BibitemShut
  {NoStop}%
\bibitem [{\citenamefont {Passarelli}\ \emph {et~al.}(2018)\citenamefont
  {Passarelli}, \citenamefont {De~Filippis}, \citenamefont {Cataudella},\ and\
  \citenamefont {Lucignano}}]{gpassarelli:qa4}%
  \BibitemOpen
  \bibfield  {author} {\bibinfo {author} {\bibfnamefont {G.}~\bibnamefont
  {Passarelli}}, \bibinfo {author} {\bibfnamefont {G.}~\bibnamefont
  {De~Filippis}}, \bibinfo {author} {\bibfnamefont {V.}~\bibnamefont
  {Cataudella}},\ and\ \bibinfo {author} {\bibfnamefont {P.}~\bibnamefont
  {Lucignano}},\ }\href {https://doi.org/10.1103/PhysRevA.97.022319} {\bibfield
   {journal} {\bibinfo  {journal} {Phys. Rev. A}\ }\textbf {\bibinfo {volume}
  {97}},\ \bibinfo {pages} {022319} (\bibinfo {year} {2018})}\BibitemShut
  {NoStop}%
\bibitem [{\citenamefont {Passarelli}\ \emph {et~al.}(2019)\citenamefont
  {Passarelli}, \citenamefont {Cataudella},\ and\ \citenamefont
  {Lucignano}}]{gpassarelli:qa1}%
  \BibitemOpen
  \bibfield  {author} {\bibinfo {author} {\bibfnamefont {G.}~\bibnamefont
  {Passarelli}}, \bibinfo {author} {\bibfnamefont {V.}~\bibnamefont
  {Cataudella}},\ and\ \bibinfo {author} {\bibfnamefont {P.}~\bibnamefont
  {Lucignano}},\ }\href {https://doi.org/10.1103/PhysRevB.100.024302}
  {\bibfield  {journal} {\bibinfo  {journal} {Phys. Rev. B}\ }\textbf {\bibinfo
  {volume} {100}},\ \bibinfo {pages} {024302} (\bibinfo {year}
  {2019})}\BibitemShut {NoStop}%
\bibitem [{\citenamefont {Marshall}\ \emph {et~al.}(2019)\citenamefont
  {Marshall}, \citenamefont {Venturelli}, \citenamefont {Hen},\ and\
  \citenamefont {Rieffel}}]{marshall:pausing}%
  \BibitemOpen
  \bibfield  {author} {\bibinfo {author} {\bibfnamefont {J.}~\bibnamefont
  {Marshall}}, \bibinfo {author} {\bibfnamefont {D.}~\bibnamefont
  {Venturelli}}, \bibinfo {author} {\bibfnamefont {I.}~\bibnamefont {Hen}},\
  and\ \bibinfo {author} {\bibfnamefont {E.~G.}\ \bibnamefont {Rieffel}},\
  }\href {https://doi.org/10.1103/PhysRevApplied.11.044083} {\bibfield
  {journal} {\bibinfo  {journal} {Phys. Rev. Applied}\ }\textbf {\bibinfo
  {volume} {11}},\ \bibinfo {pages} {044083} (\bibinfo {year}
  {2019})}\BibitemShut {NoStop}%
\bibitem [{\citenamefont {Gonzalez~Izquierdo}\ \emph
  {et~al.}(2021)\citenamefont {Gonzalez~Izquierdo}, \citenamefont {Grabbe},
  \citenamefont {Hadfield}, \citenamefont {Marshall}, \citenamefont {Wang},\
  and\ \citenamefont {Rieffel}}]{marshal:pausing2}%
  \BibitemOpen
  \bibfield  {author} {\bibinfo {author} {\bibfnamefont {Z.}~\bibnamefont
  {Gonzalez~Izquierdo}}, \bibinfo {author} {\bibfnamefont {S.}~\bibnamefont
  {Grabbe}}, \bibinfo {author} {\bibfnamefont {S.}~\bibnamefont {Hadfield}},
  \bibinfo {author} {\bibfnamefont {J.}~\bibnamefont {Marshall}}, \bibinfo
  {author} {\bibfnamefont {Z.}~\bibnamefont {Wang}},\ and\ \bibinfo {author}
  {\bibfnamefont {E.}~\bibnamefont {Rieffel}},\ }\href
  {https://doi.org/10.1103/PhysRevApplied.15.044013} {\bibfield  {journal}
  {\bibinfo  {journal} {Phys. Rev. Applied}\ }\textbf {\bibinfo {volume}
  {15}},\ \bibinfo {pages} {044013} (\bibinfo {year} {2021})}\BibitemShut
  {NoStop}%
\bibitem [{\citenamefont {Chen}\ and\ \citenamefont
  {Lidar}(2020)}]{lidar:pausing}%
  \BibitemOpen
  \bibfield  {author} {\bibinfo {author} {\bibfnamefont {H.}~\bibnamefont
  {Chen}}\ and\ \bibinfo {author} {\bibfnamefont {D.~A.}\ \bibnamefont
  {Lidar}},\ }\href {https://doi.org/10.1103/PhysRevApplied.14.014100}
  {\bibfield  {journal} {\bibinfo  {journal} {Phys. Rev. Applied}\ }\textbf
  {\bibinfo {volume} {14}},\ \bibinfo {pages} {014100} (\bibinfo {year}
  {2020})}\BibitemShut {NoStop}%
\bibitem [{\citenamefont {Passarelli}\ \emph
  {et~al.}(2020{\natexlab{a}})\citenamefont {Passarelli}, \citenamefont {Yip},
  \citenamefont {Lidar}, \citenamefont {Nishimori},\ and\ \citenamefont
  {Lucignano}}]{gpassarelli:qa2}%
  \BibitemOpen
  \bibfield  {author} {\bibinfo {author} {\bibfnamefont {G.}~\bibnamefont
  {Passarelli}}, \bibinfo {author} {\bibfnamefont {K.-W.}\ \bibnamefont {Yip}},
  \bibinfo {author} {\bibfnamefont {D.~A.}\ \bibnamefont {Lidar}}, \bibinfo
  {author} {\bibfnamefont {H.}~\bibnamefont {Nishimori}},\ and\ \bibinfo
  {author} {\bibfnamefont {P.}~\bibnamefont {Lucignano}},\ }\href
  {https://doi.org/10.1103/PhysRevA.101.022331} {\bibfield  {journal} {\bibinfo
   {journal} {Phys. Rev. A}\ }\textbf {\bibinfo {volume} {101}},\ \bibinfo
  {pages} {022331} (\bibinfo {year} {2020}{\natexlab{a}})}\BibitemShut
  {NoStop}%
\bibitem [{\citenamefont {Passarelli}\ \emph
  {et~al.}(2022{\natexlab{a}})\citenamefont {Passarelli}, \citenamefont {Yip},
  \citenamefont {Lidar},\ and\ \citenamefont {Lucignano}}]{gpassarelli:qa6}%
  \BibitemOpen
  \bibfield  {author} {\bibinfo {author} {\bibfnamefont {G.}~\bibnamefont
  {Passarelli}}, \bibinfo {author} {\bibfnamefont {K.-W.}\ \bibnamefont {Yip}},
  \bibinfo {author} {\bibfnamefont {D.~A.}\ \bibnamefont {Lidar}},\ and\
  \bibinfo {author} {\bibfnamefont {P.}~\bibnamefont {Lucignano}},\ }\href
  {https://doi.org/10.1103/PhysRevA.105.032431} {\bibfield  {journal} {\bibinfo
   {journal} {Phys. Rev. A}\ }\textbf {\bibinfo {volume} {105}},\ \bibinfo
  {pages} {032431} (\bibinfo {year} {2022}{\natexlab{a}})}\BibitemShut
  {NoStop}%
\bibitem [{\citenamefont {Ohkuwa}\ \emph {et~al.}(2018)\citenamefont {Ohkuwa},
  \citenamefont {Nishimori},\ and\ \citenamefont {Lidar}}]{ohkuwa:reverse}%
  \BibitemOpen
  \bibfield  {author} {\bibinfo {author} {\bibfnamefont {M.}~\bibnamefont
  {Ohkuwa}}, \bibinfo {author} {\bibfnamefont {H.}~\bibnamefont {Nishimori}},\
  and\ \bibinfo {author} {\bibfnamefont {D.~A.}\ \bibnamefont {Lidar}},\ }\href
  {https://doi.org/10.1103/PhysRevA.98.022314} {\bibfield  {journal} {\bibinfo
  {journal} {Phys. Rev. A}\ }\textbf {\bibinfo {volume} {98}},\ \bibinfo
  {pages} {022314} (\bibinfo {year} {2018})}\BibitemShut {NoStop}%
\bibitem [{\citenamefont {Yamashiro}\ \emph {et~al.}(2019)\citenamefont
  {Yamashiro}, \citenamefont {Ohkuwa}, \citenamefont {Nishimori},\ and\
  \citenamefont {Lidar}}]{lidar:reverse}%
  \BibitemOpen
  \bibfield  {author} {\bibinfo {author} {\bibfnamefont {Y.}~\bibnamefont
  {Yamashiro}}, \bibinfo {author} {\bibfnamefont {M.}~\bibnamefont {Ohkuwa}},
  \bibinfo {author} {\bibfnamefont {H.}~\bibnamefont {Nishimori}},\ and\
  \bibinfo {author} {\bibfnamefont {D.~A.}\ \bibnamefont {Lidar}},\ }\href
  {https://doi.org/10.1103/PhysRevA.100.052321} {\bibfield  {journal} {\bibinfo
   {journal} {Phys. Rev. A}\ }\textbf {\bibinfo {volume} {100}},\ \bibinfo
  {pages} {052321} (\bibinfo {year} {2019})}\BibitemShut {NoStop}%
\bibitem [{\citenamefont {Venturelli}\ and\ \citenamefont
  {Kondratyev}(2019{\natexlab{b}})}]{venturelli:reverse}%
  \BibitemOpen
  \bibfield  {author} {\bibinfo {author} {\bibfnamefont {D.}~\bibnamefont
  {Venturelli}}\ and\ \bibinfo {author} {\bibfnamefont {A.}~\bibnamefont
  {Kondratyev}},\ }\href {https://doi.org/10.1007/s42484-019-00001-w}
  {\bibfield  {journal} {\bibinfo  {journal} {Quantum Machine Intelligence}\
  }\textbf {\bibinfo {volume} {1}},\ \bibinfo {pages} {17} (\bibinfo {year}
  {2019}{\natexlab{b}})}\BibitemShut {NoStop}%
\bibitem [{\citenamefont {Claeys}\ \emph {et~al.}(2019)\citenamefont {Claeys},
  \citenamefont {Pandey}, \citenamefont {Sels},\ and\ \citenamefont
  {Polkovnikov}}]{claeys:variational-sta}%
  \BibitemOpen
  \bibfield  {author} {\bibinfo {author} {\bibfnamefont {P.~W.}\ \bibnamefont
  {Claeys}}, \bibinfo {author} {\bibfnamefont {M.}~\bibnamefont {Pandey}},
  \bibinfo {author} {\bibfnamefont {D.}~\bibnamefont {Sels}},\ and\ \bibinfo
  {author} {\bibfnamefont {A.}~\bibnamefont {Polkovnikov}},\ }\href
  {https://doi.org/10.1103/PhysRevLett.123.090602} {\bibfield  {journal}
  {\bibinfo  {journal} {Phys. Rev. Lett.}\ }\textbf {\bibinfo {volume} {123}},\
  \bibinfo {pages} {090602} (\bibinfo {year} {2019})}\BibitemShut {NoStop}%
\bibitem [{\citenamefont {del Campo}(2013)}]{delcampo:sta}%
  \BibitemOpen
  \bibfield  {author} {\bibinfo {author} {\bibfnamefont {A.}~\bibnamefont {del
  Campo}},\ }\href {https://doi.org/10.1103/PhysRevLett.111.100502} {\bibfield
  {journal} {\bibinfo  {journal} {Phys. Rev. Lett.}\ }\textbf {\bibinfo
  {volume} {111}},\ \bibinfo {pages} {100502} (\bibinfo {year}
  {2013})}\BibitemShut {NoStop}%
\bibitem [{\citenamefont {Gu\'ery-Odelin}\ \emph {et~al.}(2019)\citenamefont
  {Gu\'ery-Odelin}, \citenamefont {Ruschhaupt}, \citenamefont {Kiely},
  \citenamefont {Torrontegui}, \citenamefont {Mart\'{\i}nez-Garaot},\ and\
  \citenamefont {Muga}}]{guery:sta}%
  \BibitemOpen
  \bibfield  {author} {\bibinfo {author} {\bibfnamefont {D.}~\bibnamefont
  {Gu\'ery-Odelin}}, \bibinfo {author} {\bibfnamefont {A.}~\bibnamefont
  {Ruschhaupt}}, \bibinfo {author} {\bibfnamefont {A.}~\bibnamefont {Kiely}},
  \bibinfo {author} {\bibfnamefont {E.}~\bibnamefont {Torrontegui}}, \bibinfo
  {author} {\bibfnamefont {S.}~\bibnamefont {Mart\'{\i}nez-Garaot}},\ and\
  \bibinfo {author} {\bibfnamefont {J.~G.}\ \bibnamefont {Muga}},\ }\href
  {https://doi.org/10.1103/RevModPhys.91.045001} {\bibfield  {journal}
  {\bibinfo  {journal} {Rev. Mod. Phys.}\ }\textbf {\bibinfo {volume} {91}},\
  \bibinfo {pages} {045001} (\bibinfo {year} {2019})}\BibitemShut {NoStop}%
\bibitem [{\citenamefont {Sels}\ and\ \citenamefont
  {Polkovnikov}(2017)}]{sels:sta}%
  \BibitemOpen
  \bibfield  {author} {\bibinfo {author} {\bibfnamefont {D.}~\bibnamefont
  {Sels}}\ and\ \bibinfo {author} {\bibfnamefont {A.}~\bibnamefont
  {Polkovnikov}},\ }\bibfield  {journal} {\bibinfo  {journal} {Proceedings of
  the National Academy of Sciences}\ }\textbf {\bibinfo {volume} {114}},\ \href
  {https://doi.org/10.1073/pnas.1619826114} {10.1073/pnas.1619826114} (\bibinfo
  {year} {2017})\BibitemShut {NoStop}%
\bibitem [{\citenamefont {Passarelli}\ \emph
  {et~al.}(2020{\natexlab{b}})\citenamefont {Passarelli}, \citenamefont
  {Cataudella}, \citenamefont {Fazio},\ and\ \citenamefont
  {Lucignano}}]{gpassarelli:qa3}%
  \BibitemOpen
  \bibfield  {author} {\bibinfo {author} {\bibfnamefont {G.}~\bibnamefont
  {Passarelli}}, \bibinfo {author} {\bibfnamefont {V.}~\bibnamefont
  {Cataudella}}, \bibinfo {author} {\bibfnamefont {R.}~\bibnamefont {Fazio}},\
  and\ \bibinfo {author} {\bibfnamefont {P.}~\bibnamefont {Lucignano}},\ }\href
  {https://doi.org/10.1103/PhysRevResearch.2.013283} {\bibfield  {journal}
  {\bibinfo  {journal} {Phys. Rev. Research}\ }\textbf {\bibinfo {volume}
  {2}},\ \bibinfo {pages} {013283} (\bibinfo {year}
  {2020}{\natexlab{b}})}\BibitemShut {NoStop}%
\bibitem [{\citenamefont {Torrontegui}\ \emph {et~al.}(2013)\citenamefont
  {Torrontegui}, \citenamefont {Ibáñez}, \citenamefont {Martínez-Garaot},
  \citenamefont {Modugno}, \citenamefont {{del Campo}}, \citenamefont
  {Guéry-Odelin}, \citenamefont {Ruschhaupt}, \citenamefont {Chen},\ and\
  \citenamefont {Muga}}]{TORRONTEGUI:sta}%
  \BibitemOpen
  \bibfield  {author} {\bibinfo {author} {\bibfnamefont {E.}~\bibnamefont
  {Torrontegui}}, \bibinfo {author} {\bibfnamefont {S.}~\bibnamefont
  {Ibáñez}}, \bibinfo {author} {\bibfnamefont {S.}~\bibnamefont
  {Martínez-Garaot}}, \bibinfo {author} {\bibfnamefont {M.}~\bibnamefont
  {Modugno}}, \bibinfo {author} {\bibfnamefont {A.}~\bibnamefont {{del
  Campo}}}, \bibinfo {author} {\bibfnamefont {D.}~\bibnamefont
  {Guéry-Odelin}}, \bibinfo {author} {\bibfnamefont {A.}~\bibnamefont
  {Ruschhaupt}}, \bibinfo {author} {\bibfnamefont {X.}~\bibnamefont {Chen}},\
  and\ \bibinfo {author} {\bibfnamefont {J.~G.}\ \bibnamefont {Muga}},\ }in\
  \href {https://doi.org/https://doi.org/10.1016/B978-0-12-408090-4.00002-5}
  {\emph {\bibinfo {booktitle} {Advances in Atomic, Molecular, and Optical
  Physics}}},\ \bibinfo {series} {Advances In Atomic, Molecular, and Optical
  Physics}, Vol.~\bibinfo {volume} {62},\ \bibinfo {editor} {edited by\
  \bibinfo {editor} {\bibfnamefont {E.}~\bibnamefont {Arimondo}}, \bibinfo
  {editor} {\bibfnamefont {P.~R.}\ \bibnamefont {Berman}},\ and\ \bibinfo
  {editor} {\bibfnamefont {C.~C.}\ \bibnamefont {Lin}}}\ (\bibinfo  {publisher}
  {Academic Press},\ \bibinfo {year} {2013})\ pp.\ \bibinfo {pages}
  {117--169}\BibitemShut {NoStop}%
\bibitem [{\citenamefont {Passarelli}\ \emph
  {et~al.}(2022{\natexlab{b}})\citenamefont {Passarelli}, \citenamefont
  {Fazio},\ and\ \citenamefont {Lucignano}}]{gpassarelli:qa5}%
  \BibitemOpen
  \bibfield  {author} {\bibinfo {author} {\bibfnamefont {G.}~\bibnamefont
  {Passarelli}}, \bibinfo {author} {\bibfnamefont {R.}~\bibnamefont {Fazio}},\
  and\ \bibinfo {author} {\bibfnamefont {P.}~\bibnamefont {Lucignano}},\ }\href
  {https://doi.org/10.1103/PhysRevA.105.022618} {\bibfield  {journal} {\bibinfo
   {journal} {Phys. Rev. A}\ }\textbf {\bibinfo {volume} {105}},\ \bibinfo
  {pages} {022618} (\bibinfo {year} {2022}{\natexlab{b}})}\BibitemShut
  {NoStop}%
\bibitem [{\citenamefont {Crosson}\ \emph {et~al.}(2014)\citenamefont
  {Crosson}, \citenamefont {Farhi}, \citenamefont {Lin}, \citenamefont {Lin},\
  and\ \citenamefont {Shor}}]{crosson:sta}%
  \BibitemOpen
  \bibfield  {author} {\bibinfo {author} {\bibfnamefont {E.}~\bibnamefont
  {Crosson}}, \bibinfo {author} {\bibfnamefont {E.}~\bibnamefont {Farhi}},
  \bibinfo {author} {\bibfnamefont {C.~Y.-Y.}\ \bibnamefont {Lin}}, \bibinfo
  {author} {\bibfnamefont {H.-H.}\ \bibnamefont {Lin}},\ and\ \bibinfo {author}
  {\bibfnamefont {P.}~\bibnamefont {Shor}},\ }\href
  {https://doi.org/10.48550/ARXIV.1401.7320} {\bibinfo {title} {Different
  strategies for optimization using the quantum adiabatic algorithm}} (\bibinfo
  {year} {2014})\BibitemShut {NoStop}%
\bibitem [{\citenamefont {Seki}\ and\ \citenamefont
  {Nishimori}(2012)}]{seki:sta}%
  \BibitemOpen
  \bibfield  {author} {\bibinfo {author} {\bibfnamefont {Y.}~\bibnamefont
  {Seki}}\ and\ \bibinfo {author} {\bibfnamefont {H.}~\bibnamefont
  {Nishimori}},\ }\href {https://doi.org/10.1103/PhysRevE.85.051112} {\bibfield
   {journal} {\bibinfo  {journal} {Phys. Rev. E}\ }\textbf {\bibinfo {volume}
  {85}},\ \bibinfo {pages} {051112} (\bibinfo {year} {2012})}\BibitemShut
  {NoStop}%
\bibitem [{\citenamefont {Somma}\ \emph {et~al.}(2012)\citenamefont {Somma},
  \citenamefont {Nagaj},\ and\ \citenamefont {Kieferov{\'a}}}]{Somma:2012kx}%
  \BibitemOpen
  \bibfield  {author} {\bibinfo {author} {\bibfnamefont {R.~D.}\ \bibnamefont
  {Somma}}, \bibinfo {author} {\bibfnamefont {D.}~\bibnamefont {Nagaj}},\ and\
  \bibinfo {author} {\bibfnamefont {M.}~\bibnamefont {Kieferov{\'a}}},\ }\href
  {http://link.aps.org/doi/10.1103/PhysRevLett.109.050501} {\bibfield
  {journal} {\bibinfo  {journal} {Phys. Rev. Lett.}\ }\textbf {\bibinfo
  {volume} {109}},\ \bibinfo {pages} {050501} (\bibinfo {year}
  {2012})}\BibitemShut {NoStop}%
\bibitem [{\citenamefont {Brady}\ \emph {et~al.}(2021)\citenamefont {Brady},
  \citenamefont {Baldwin}, \citenamefont {Bapat}, \citenamefont {Kharkov},\
  and\ \citenamefont {Gorshkov}}]{brady_optimal_2021}%
  \BibitemOpen
  \bibfield  {author} {\bibinfo {author} {\bibfnamefont {L.~T.}\ \bibnamefont
  {Brady}}, \bibinfo {author} {\bibfnamefont {C.~L.}\ \bibnamefont {Baldwin}},
  \bibinfo {author} {\bibfnamefont {A.}~\bibnamefont {Bapat}}, \bibinfo
  {author} {\bibfnamefont {Y.}~\bibnamefont {Kharkov}},\ and\ \bibinfo {author}
  {\bibfnamefont {A.~V.}\ \bibnamefont {Gorshkov}},\ }\href
  {https://doi.org/10.1103/PhysRevLett.126.070505} {\bibfield  {journal}
  {\bibinfo  {journal} {Phys. Rev. Lett.}\ }\textbf {\bibinfo {volume} {126}},\
  \bibinfo {pages} {070505} (\bibinfo {year} {2021})}\BibitemShut {NoStop}%
\bibitem [{\citenamefont {Venuti}\ \emph {et~al.}(2021)\citenamefont {Venuti},
  \citenamefont {D'Alessandro},\ and\ \citenamefont
  {Lidar}}]{venuti2021optimal}%
  \BibitemOpen
  \bibfield  {author} {\bibinfo {author} {\bibfnamefont {L.~C.}\ \bibnamefont
  {Venuti}}, \bibinfo {author} {\bibfnamefont {D.}~\bibnamefont
  {D'Alessandro}},\ and\ \bibinfo {author} {\bibfnamefont {D.~A.}\ \bibnamefont
  {Lidar}},\ }\href {https://doi.org/10.1103/PhysRevApplied.16.054023}
  {\bibfield  {journal} {\bibinfo  {journal} {Physical Review Applied}\
  }\textbf {\bibinfo {volume} {16}},\ \bibinfo {pages} {054023} (\bibinfo
  {year} {2021})}\BibitemShut {NoStop}%
\bibitem [{\citenamefont {Crosson}\ and\ \citenamefont
  {Lidar}(2021)}]{Crosson2021}%
  \BibitemOpen
  \bibfield  {author} {\bibinfo {author} {\bibfnamefont {E.~J.}\ \bibnamefont
  {Crosson}}\ and\ \bibinfo {author} {\bibfnamefont {D.~A.}\ \bibnamefont
  {Lidar}},\ }\href {https://doi.org/10.1038/s42254-021-00313-6} {\bibfield
  {journal} {\bibinfo  {journal} {Nature Reviews Physics}\ }\textbf {\bibinfo
  {volume} {3}},\ \bibinfo {pages} {466} (\bibinfo {year} {2021})}\BibitemShut
  {NoStop}%
\bibitem [{\citenamefont {Farhi}\ \emph {et~al.}(2010)\citenamefont {Farhi},
  \citenamefont {Goldstone}, \citenamefont {Gosset}, \citenamefont {Gutmann},
  \citenamefont {Meyer},\ and\ \citenamefont {Shor}}]{farhi:sta}%
  \BibitemOpen
  \bibfield  {author} {\bibinfo {author} {\bibfnamefont {E.}~\bibnamefont
  {Farhi}}, \bibinfo {author} {\bibfnamefont {J.}~\bibnamefont {Goldstone}},
  \bibinfo {author} {\bibfnamefont {D.}~\bibnamefont {Gosset}}, \bibinfo
  {author} {\bibfnamefont {S.}~\bibnamefont {Gutmann}}, \bibinfo {author}
  {\bibfnamefont {H.~B.}\ \bibnamefont {Meyer}},\ and\ \bibinfo {author}
  {\bibfnamefont {P.}~\bibnamefont {Shor}},\ }\href
  {http://arxiv.org/abs/0909.4766} {\bibinfo {title} {Quantum {Adiabatic}
  {Algorithms}, {Small} {Gaps}, and {Different} {Paths}}} (\bibinfo {year}
  {2010}),\ \bibinfo {note} {arXiv:0909.4766 [quant-ph]}\BibitemShut {NoStop}%
\bibitem [{\citenamefont {Côté}\ \emph {et~al.}(2022)\citenamefont {Côté},
  \citenamefont {Sauvage}, \citenamefont {Larocca}, \citenamefont {Jonsson},
  \citenamefont {Cincio},\ and\ \citenamefont {Albash}}]{cote-diabatic}%
  \BibitemOpen
  \bibfield  {author} {\bibinfo {author} {\bibfnamefont {J.}~\bibnamefont
  {Côté}}, \bibinfo {author} {\bibfnamefont {F.}~\bibnamefont {Sauvage}},
  \bibinfo {author} {\bibfnamefont {M.}~\bibnamefont {Larocca}}, \bibinfo
  {author} {\bibfnamefont {M.}~\bibnamefont {Jonsson}}, \bibinfo {author}
  {\bibfnamefont {L.}~\bibnamefont {Cincio}},\ and\ \bibinfo {author}
  {\bibfnamefont {T.}~\bibnamefont {Albash}},\ }\href@noop {} {\bibinfo {title}
  {Diabatic quantum annealing for the frustrated ring model}} (\bibinfo {year}
  {2022}),\ \Eprint {https://arxiv.org/abs/2212.02624} {arXiv:2212.02624
  [quant-ph]} \BibitemShut {NoStop}%
\bibitem [{\citenamefont {Durkin}(2019)}]{durkin:catalystH}%
  \BibitemOpen
  \bibfield  {author} {\bibinfo {author} {\bibfnamefont {G.~A.}\ \bibnamefont
  {Durkin}},\ }\href {https://doi.org/10.1103/PhysRevA.99.032315} {\bibfield
  {journal} {\bibinfo  {journal} {Phys. Rev. A}\ }\textbf {\bibinfo {volume}
  {99}},\ \bibinfo {pages} {032315} (\bibinfo {year} {2019})}\BibitemShut
  {NoStop}%
\bibitem [{\citenamefont {Albash}(2019)}]{albash:catalystH}%
  \BibitemOpen
  \bibfield  {author} {\bibinfo {author} {\bibfnamefont {T.}~\bibnamefont
  {Albash}},\ }\href {https://doi.org/10.1103/PhysRevA.99.042334} {\bibfield
  {journal} {\bibinfo  {journal} {Phys. Rev. A}\ }\textbf {\bibinfo {volume}
  {99}},\ \bibinfo {pages} {042334} (\bibinfo {year} {2019})}\BibitemShut
  {NoStop}%
\bibitem [{\citenamefont {Roland}\ and\ \citenamefont
  {Cerf}(2002)}]{local-adiab:roland-cerf}%
  \BibitemOpen
  \bibfield  {author} {\bibinfo {author} {\bibfnamefont {J.}~\bibnamefont
  {Roland}}\ and\ \bibinfo {author} {\bibfnamefont {N.~J.}\ \bibnamefont
  {Cerf}},\ }\href {https://doi.org/10.1103/PhysRevA.65.042308} {\bibfield
  {journal} {\bibinfo  {journal} {Phys. Rev. A}\ }\textbf {\bibinfo {volume}
  {65}},\ \bibinfo {pages} {042308} (\bibinfo {year} {2002})}\BibitemShut
  {NoStop}%
\bibitem [{\citenamefont {Khezri}\ \emph {et~al.}(2022)\citenamefont {Khezri},
  \citenamefont {Dai}, \citenamefont {Yang}, \citenamefont {Albash},
  \citenamefont {Lupascu},\ and\ \citenamefont {Lidar}}]{khezri:qa-sch}%
  \BibitemOpen
  \bibfield  {author} {\bibinfo {author} {\bibfnamefont {M.}~\bibnamefont
  {Khezri}}, \bibinfo {author} {\bibfnamefont {X.}~\bibnamefont {Dai}},
  \bibinfo {author} {\bibfnamefont {R.}~\bibnamefont {Yang}}, \bibinfo {author}
  {\bibfnamefont {T.}~\bibnamefont {Albash}}, \bibinfo {author} {\bibfnamefont
  {A.}~\bibnamefont {Lupascu}},\ and\ \bibinfo {author} {\bibfnamefont {D.~A.}\
  \bibnamefont {Lidar}},\ }\href
  {https://doi.org/10.1103/PhysRevApplied.17.044005} {\bibfield  {journal}
  {\bibinfo  {journal} {Phys. Rev. Applied}\ }\textbf {\bibinfo {volume}
  {17}},\ \bibinfo {pages} {044005} (\bibinfo {year} {2022})}\BibitemShut
  {NoStop}%
\bibitem [{\citenamefont {Morita}\ and\ \citenamefont
  {Nishimori}(2008)}]{morita:qa-sch}%
  \BibitemOpen
  \bibfield  {author} {\bibinfo {author} {\bibfnamefont {S.}~\bibnamefont
  {Morita}}\ and\ \bibinfo {author} {\bibfnamefont {H.}~\bibnamefont
  {Nishimori}},\ }\href {https://doi.org/10.1063/1.2995837} {\bibfield
  {journal} {\bibinfo  {journal} {Journal of Mathematical Physics}\ }\textbf
  {\bibinfo {volume} {49}},\ \bibinfo {pages} {125210} (\bibinfo {year}
  {2008})}\BibitemShut {NoStop}%
\bibitem [{\citenamefont {Matsuura}\ \emph {et~al.}(2021)\citenamefont
  {Matsuura}, \citenamefont {Buck}, \citenamefont {Senicourt},\ and\
  \citenamefont {Zaribafiyan}}]{matsuura-timesch-optimization}%
  \BibitemOpen
  \bibfield  {author} {\bibinfo {author} {\bibfnamefont {S.}~\bibnamefont
  {Matsuura}}, \bibinfo {author} {\bibfnamefont {S.}~\bibnamefont {Buck}},
  \bibinfo {author} {\bibfnamefont {V.}~\bibnamefont {Senicourt}},\ and\
  \bibinfo {author} {\bibfnamefont {A.}~\bibnamefont {Zaribafiyan}},\ }\href
  {https://doi.org/10.1103/PhysRevA.103.052435} {\bibfield  {journal} {\bibinfo
   {journal} {Phys. Rev. A}\ }\textbf {\bibinfo {volume} {103}},\ \bibinfo
  {pages} {052435} (\bibinfo {year} {2021})}\BibitemShut {NoStop}%
\bibitem [{\citenamefont {Susa}\ and\ \citenamefont
  {Nishimori}(2021)}]{susa-nishimori-timesch-optimization}%
  \BibitemOpen
  \bibfield  {author} {\bibinfo {author} {\bibfnamefont {Y.}~\bibnamefont
  {Susa}}\ and\ \bibinfo {author} {\bibfnamefont {H.}~\bibnamefont
  {Nishimori}},\ }\href {https://doi.org/10.1103/PhysRevA.103.022619}
  {\bibfield  {journal} {\bibinfo  {journal} {Phys. Rev. A}\ }\textbf {\bibinfo
  {volume} {103}},\ \bibinfo {pages} {022619} (\bibinfo {year}
  {2021})}\BibitemShut {NoStop}%
\bibitem [{\citenamefont {Chen}\ \emph {et~al.}(2022)\citenamefont {Chen},
  \citenamefont {Chen}, \citenamefont {Lee}, \citenamefont {Zhang},\ and\
  \citenamefont {Hsieh}}]{chen:sch-ml-3sat}%
  \BibitemOpen
  \bibfield  {author} {\bibinfo {author} {\bibfnamefont {Y.-Q.}\ \bibnamefont
  {Chen}}, \bibinfo {author} {\bibfnamefont {Y.}~\bibnamefont {Chen}}, \bibinfo
  {author} {\bibfnamefont {C.-K.}\ \bibnamefont {Lee}}, \bibinfo {author}
  {\bibfnamefont {S.}~\bibnamefont {Zhang}},\ and\ \bibinfo {author}
  {\bibfnamefont {C.-Y.}\ \bibnamefont {Hsieh}},\ }\href
  {https://doi.org/10.1038/s42256-022-00446-y} {\bibfield  {journal} {\bibinfo
  {journal} {Nature Machine Intelligence}\ }\textbf {\bibinfo {volume} {4}},\
  \bibinfo {pages} {269} (\bibinfo {year} {2022})}\BibitemShut {NoStop}%
\bibitem [{\citenamefont {Hegde}\ \emph {et~al.}(2022)\citenamefont {Hegde},
  \citenamefont {Passarelli}, \citenamefont {Scocco},\ and\ \citenamefont
  {Lucignano}}]{phegde:ga}%
  \BibitemOpen
  \bibfield  {author} {\bibinfo {author} {\bibfnamefont {P.~R.}\ \bibnamefont
  {Hegde}}, \bibinfo {author} {\bibfnamefont {G.}~\bibnamefont {Passarelli}},
  \bibinfo {author} {\bibfnamefont {A.}~\bibnamefont {Scocco}},\ and\ \bibinfo
  {author} {\bibfnamefont {P.}~\bibnamefont {Lucignano}},\ }\href
  {https://doi.org/10.1103/PhysRevA.105.012612} {\bibfield  {journal} {\bibinfo
   {journal} {Phys. Rev. A}\ }\textbf {\bibinfo {volume} {105}},\ \bibinfo
  {pages} {012612} (\bibinfo {year} {2022})}\BibitemShut {NoStop}%
\bibitem [{\citenamefont {Choi}(2008)}]{choi_minor-embedding_2008}%
  \BibitemOpen
  \bibfield  {author} {\bibinfo {author} {\bibfnamefont {V.}~\bibnamefont
  {Choi}},\ }\href {https://doi.org/10.1007/s11128-008-0082-9} {\bibfield
  {journal} {\bibinfo  {journal} {Quantum Information Processing}\ }\textbf
  {\bibinfo {volume} {7}},\ \bibinfo {pages} {193} (\bibinfo {year}
  {2008})}\BibitemShut {NoStop}%
\bibitem [{\citenamefont {Acampora}\ \emph {et~al.}(2019)\citenamefont
  {Acampora}, \citenamefont {Cataudella}, \citenamefont {Hegde}, \citenamefont
  {Lucignano}, \citenamefont {Passarelli},\ and\ \citenamefont
  {Vitiello}}]{acampora:genetic}%
  \BibitemOpen
  \bibfield  {author} {\bibinfo {author} {\bibfnamefont {G.}~\bibnamefont
  {Acampora}}, \bibinfo {author} {\bibfnamefont {V.}~\bibnamefont
  {Cataudella}}, \bibinfo {author} {\bibfnamefont {P.~R.}\ \bibnamefont
  {Hegde}}, \bibinfo {author} {\bibfnamefont {P.}~\bibnamefont {Lucignano}},
  \bibinfo {author} {\bibfnamefont {G.}~\bibnamefont {Passarelli}},\ and\
  \bibinfo {author} {\bibfnamefont {A.}~\bibnamefont {Vitiello}},\ }\href
  {https://doi.org/https://doi.org/10.1007/s42484-019-00011-8} {\bibfield
  {journal} {\bibinfo  {journal} {Quantum Machine Intelligence}\ }\textbf
  {\bibinfo {volume} {1}},\ \bibinfo {pages} {113} (\bibinfo {year}
  {2019})}\BibitemShut {NoStop}%
\bibitem [{\citenamefont {Acampora}\ \emph {et~al.}(2021)\citenamefont
  {Acampora}, \citenamefont {Cataudella}, \citenamefont {Hegde}, \citenamefont
  {Lucignano}, \citenamefont {Passarelli},\ and\ \citenamefont
  {Vitiello}}]{acampora:genetic2}%
  \BibitemOpen
  \bibfield  {author} {\bibinfo {author} {\bibfnamefont {G.}~\bibnamefont
  {Acampora}}, \bibinfo {author} {\bibfnamefont {V.}~\bibnamefont
  {Cataudella}}, \bibinfo {author} {\bibfnamefont {P.~R.}\ \bibnamefont
  {Hegde}}, \bibinfo {author} {\bibfnamefont {P.}~\bibnamefont {Lucignano}},
  \bibinfo {author} {\bibfnamefont {G.}~\bibnamefont {Passarelli}},\ and\
  \bibinfo {author} {\bibfnamefont {A.}~\bibnamefont {Vitiello}},\ }\href
  {https://doi.org/https://doi.org/10.1016/j.asoc.2021.107634} {\bibfield
  {journal} {\bibinfo  {journal} {Applied Soft Computing}\ }\textbf {\bibinfo
  {volume} {110}},\ \bibinfo {pages} {107634} (\bibinfo {year}
  {2021})}\BibitemShut {NoStop}%
\bibitem [{\citenamefont {Zaman}\ \emph {et~al.}(2022)\citenamefont {Zaman},
  \citenamefont {Tanahashi},\ and\ \citenamefont {Tanaka}}]{zaman:pyqubo}%
  \BibitemOpen
  \bibfield  {author} {\bibinfo {author} {\bibfnamefont {M.}~\bibnamefont
  {Zaman}}, \bibinfo {author} {\bibfnamefont {K.}~\bibnamefont {Tanahashi}},\
  and\ \bibinfo {author} {\bibfnamefont {S.}~\bibnamefont {Tanaka}},\ }\href
  {https://doi.org/10.1109/TC.2021.3063618} {\bibfield  {journal} {\bibinfo
  {journal} {IEEE Transactions on Computers}\ }\textbf {\bibinfo {volume}
  {71}},\ \bibinfo {pages} {838} (\bibinfo {year} {2022})}\BibitemShut
  {NoStop}%
\bibitem [{\citenamefont {Hen}\ and\ \citenamefont
  {Spedalieri}(2016)}]{Hen:embedding}%
  \BibitemOpen
  \bibfield  {author} {\bibinfo {author} {\bibfnamefont {I.}~\bibnamefont
  {Hen}}\ and\ \bibinfo {author} {\bibfnamefont {F.~M.}\ \bibnamefont
  {Spedalieri}},\ }\href {https://doi.org/10.1103/PhysRevApplied.5.034007}
  {\bibfield  {journal} {\bibinfo  {journal} {Phys. Rev. Appl.}\ }\textbf
  {\bibinfo {volume} {5}},\ \bibinfo {pages} {034007} (\bibinfo {year}
  {2016})}\BibitemShut {NoStop}%
\bibitem [{\citenamefont {Barahona}(1982)}]{barahona:isingmodel}%
  \BibitemOpen
  \bibfield  {author} {\bibinfo {author} {\bibfnamefont {F.}~\bibnamefont
  {Barahona}},\ }\href {https://doi.org/10.1088/0305-4470/15/10/028} {\bibfield
   {journal} {\bibinfo  {journal} {Journal of Physics A: Mathematical and
  General}\ }\textbf {\bibinfo {volume} {15}},\ \bibinfo {pages} {3241}
  (\bibinfo {year} {1982})}\BibitemShut {NoStop}%
\bibitem [{\citenamefont {Mitchell}(1997)}]{mitchell_machine_1997}%
  \BibitemOpen
  \bibfield  {author} {\bibinfo {author} {\bibfnamefont {T.~M.}\ \bibnamefont
  {Mitchell}},\ }\href@noop {} {\emph {\bibinfo {title} {Machine
  {Learning}}}},\ {McGraw}-{Hill} series in computer science\ (\bibinfo
  {publisher} {McGraw-Hill},\ \bibinfo {address} {New York},\ \bibinfo {year}
  {1997})\BibitemShut {NoStop}%
\bibitem [{\citenamefont {Mohseni}\ \emph {et~al.}(2021)\citenamefont
  {Mohseni}, \citenamefont {Navarrete-Benlloch}, \citenamefont {Byrnes},\ and\
  \citenamefont {Marquardt}}]{lstm-sch:mohseni}%
  \BibitemOpen
  \bibfield  {author} {\bibinfo {author} {\bibfnamefont {N.}~\bibnamefont
  {Mohseni}}, \bibinfo {author} {\bibfnamefont {C.}~\bibnamefont
  {Navarrete-Benlloch}}, \bibinfo {author} {\bibfnamefont {T.}~\bibnamefont
  {Byrnes}},\ and\ \bibinfo {author} {\bibfnamefont {F.}~\bibnamefont
  {Marquardt}},\ }\href {http://arxiv.org/abs/2109.08492} {\emph {\bibinfo
  {title} {Deep recurrent networks predicting the gap evolution in adiabatic
  quantum computing}}},\ \bibinfo {type} {Tech. Rep.}\ \bibinfo {number}
  {arXiv:2109.08492}\ (\bibinfo  {institution} {arXiv},\ \bibinfo {year}
  {2021})\ \bibinfo {note} {arXiv:2109.08492 [quant-ph] type:
  article}\BibitemShut {NoStop}%
\bibitem [{\citenamefont {Mohseni}\ \emph {et~al.}(2022)\citenamefont
  {Mohseni}, \citenamefont {Fösel}, \citenamefont {Guo}, \citenamefont
  {Navarrete-Benlloch},\ and\ \citenamefont {Marquardt}}]{mohseni:lstm2}%
  \BibitemOpen
  \bibfield  {author} {\bibinfo {author} {\bibfnamefont {N.}~\bibnamefont
  {Mohseni}}, \bibinfo {author} {\bibfnamefont {T.}~\bibnamefont {Fösel}},
  \bibinfo {author} {\bibfnamefont {L.}~\bibnamefont {Guo}}, \bibinfo {author}
  {\bibfnamefont {C.}~\bibnamefont {Navarrete-Benlloch}},\ and\ \bibinfo
  {author} {\bibfnamefont {F.}~\bibnamefont {Marquardt}},\ }\href
  {https://doi.org/10.22331/q-2022-05-17-714} {\bibfield  {journal} {\bibinfo
  {journal} {Quantum}\ }\textbf {\bibinfo {volume} {6}},\ \bibinfo {pages}
  {714} (\bibinfo {year} {2022})}\BibitemShut {NoStop}%
\bibitem [{\citenamefont {Grover}(1997)}]{grover}%
  \BibitemOpen
  \bibfield  {author} {\bibinfo {author} {\bibfnamefont {L.~K.}\ \bibnamefont
  {Grover}},\ }\href {https://doi.org/10.1103/PhysRevLett.79.325} {\bibfield
  {journal} {\bibinfo  {journal} {Phys. Rev. Lett.}\ }\textbf {\bibinfo
  {volume} {79}},\ \bibinfo {pages} {325} (\bibinfo {year} {1997})}\BibitemShut
  {NoStop}%
\bibitem [{\citenamefont {{Bigan Mbeng}}\ \emph {et~al.}(2019)\citenamefont
  {{Bigan Mbeng}}, \citenamefont {{Fazio}},\ and\ \citenamefont
  {{Santoro}}}]{mbeng:roland-cerf}%
  \BibitemOpen
  \bibfield  {author} {\bibinfo {author} {\bibfnamefont {G.}~\bibnamefont
  {{Bigan Mbeng}}}, \bibinfo {author} {\bibfnamefont {R.}~\bibnamefont
  {{Fazio}}},\ and\ \bibinfo {author} {\bibfnamefont {G.~E.}\ \bibnamefont
  {{Santoro}}},\ }\href@noop {} {\bibfield  {journal} {\bibinfo  {journal}
  {arXiv e-prints}\ } (\bibinfo {year} {2019})},\ \Eprint
  {https://arxiv.org/abs/1911.12259} {arXiv:1911.12259 [quant-ph]} \BibitemShut
  {NoStop}%
\bibitem [{\citenamefont {Stefanatos}\ and\ \citenamefont
  {Paspalakis}(2020)}]{Stefanatos:roland-cerf}%
  \BibitemOpen
  \bibfield  {author} {\bibinfo {author} {\bibfnamefont {D.}~\bibnamefont
  {Stefanatos}}\ and\ \bibinfo {author} {\bibfnamefont {E.}~\bibnamefont
  {Paspalakis}},\ }\href {https://doi.org/10.1088/1751-8121/ab7423} {\bibfield
  {journal} {\bibinfo  {journal} {Journal of Physics A: Mathematical and
  Theoretical}\ }\textbf {\bibinfo {volume} {53}},\ \bibinfo {pages} {115304}
  (\bibinfo {year} {2020})}\BibitemShut {NoStop}%
\bibitem [{\citenamefont {Delvecchio}\ \emph {et~al.}(2021)\citenamefont
  {Delvecchio}, \citenamefont {Petiziol},\ and\ \citenamefont
  {Wimberger}}]{Delvecchio:roland-cerf}%
  \BibitemOpen
  \bibfield  {author} {\bibinfo {author} {\bibfnamefont {M.}~\bibnamefont
  {Delvecchio}}, \bibinfo {author} {\bibfnamefont {F.}~\bibnamefont
  {Petiziol}},\ and\ \bibinfo {author} {\bibfnamefont {S.}~\bibnamefont
  {Wimberger}},\ }\bibfield  {journal} {\bibinfo  {journal} {Entropy}\ }\textbf
  {\bibinfo {volume} {23}},\ \href {https://doi.org/10.3390/e23070897}
  {10.3390/e23070897} (\bibinfo {year} {2021})\BibitemShut {NoStop}%
\bibitem [{\citenamefont {Sarjala}\ \emph {et~al.}(2006)\citenamefont
  {Sarjala}, \citenamefont {Petäjä},\ and\ \citenamefont
  {Alava}}]{Sarjala:random-qa-ising}%
  \BibitemOpen
  \bibfield  {author} {\bibinfo {author} {\bibfnamefont {M.}~\bibnamefont
  {Sarjala}}, \bibinfo {author} {\bibfnamefont {V.}~\bibnamefont {Petäjä}},\
  and\ \bibinfo {author} {\bibfnamefont {M.}~\bibnamefont {Alava}},\ }\href
  {https://doi.org/10.1088/1742-5468/2006/01/p01008} {\bibfield  {journal}
  {\bibinfo  {journal} {Journal of Statistical Mechanics: Theory and
  Experiment}\ }\textbf {\bibinfo {volume} {2006}},\ \bibinfo {pages} {P01008}
  (\bibinfo {year} {2006})}\BibitemShut {NoStop}%
\bibitem [{\citenamefont {Suzuki}\ \emph {et~al.}(2007)\citenamefont {Suzuki},
  \citenamefont {Nishimori},\ and\ \citenamefont
  {Suzuki}}]{suzuki:qa-random-ising}%
  \BibitemOpen
  \bibfield  {author} {\bibinfo {author} {\bibfnamefont {S.}~\bibnamefont
  {Suzuki}}, \bibinfo {author} {\bibfnamefont {H.}~\bibnamefont {Nishimori}},\
  and\ \bibinfo {author} {\bibfnamefont {M.}~\bibnamefont {Suzuki}},\ }\href
  {https://doi.org/10.1103/PhysRevE.75.051112} {\bibfield  {journal} {\bibinfo
  {journal} {Phys. Rev. E}\ }\textbf {\bibinfo {volume} {75}},\ \bibinfo
  {pages} {051112} (\bibinfo {year} {2007})}\BibitemShut {NoStop}%
\bibitem [{\citenamefont {Crooks}(2018)}]{crooks:qaoa}%
  \BibitemOpen
  \bibfield  {author} {\bibinfo {author} {\bibfnamefont {G.~E.}\ \bibnamefont
  {Crooks}},\ }\href {http://arxiv.org/abs/1811.08419} {\bibfield  {journal}
  {\bibinfo  {journal} {arXiv:1811.08419 [quant-ph]}\ } (\bibinfo {year}
  {2018})},\ \bibinfo {note} {arXiv: 1811.08419}\BibitemShut {NoStop}%
\bibitem [{\citenamefont {Farhi}\ \emph {et~al.}(2014)\citenamefont {Farhi},
  \citenamefont {Goldstone},\ and\ \citenamefont {Gutmann}}]{farhi:qaoa}%
  \BibitemOpen
  \bibfield  {author} {\bibinfo {author} {\bibfnamefont {E.}~\bibnamefont
  {Farhi}}, \bibinfo {author} {\bibfnamefont {J.}~\bibnamefont {Goldstone}},\
  and\ \bibinfo {author} {\bibfnamefont {S.}~\bibnamefont {Gutmann}},\ }\href
  {http://arxiv.org/abs/1411.4028} {\bibfield  {journal} {\bibinfo  {journal}
  {arXiv:1411.4028 [quant-ph]}\ } (\bibinfo {year} {2014})},\ \bibinfo {note}
  {arXiv: 1411.4028}\BibitemShut {NoStop}%
\bibitem [{\citenamefont {Johansson}\ \emph {et~al.}(2012)\citenamefont
  {Johansson}, \citenamefont {Nation},\ and\ \citenamefont {Nori}}]{qutip}%
  \BibitemOpen
  \bibfield  {author} {\bibinfo {author} {\bibfnamefont {J.}~\bibnamefont
  {Johansson}}, \bibinfo {author} {\bibfnamefont {P.}~\bibnamefont {Nation}},\
  and\ \bibinfo {author} {\bibfnamefont {F.}~\bibnamefont {Nori}},\ }\href
  {https://doi.org/https://doi.org/10.1016/j.cpc.2012.02.021} {\bibfield
  {journal} {\bibinfo  {journal} {Computer Physics Communications}\ }\textbf
  {\bibinfo {volume} {183}},\ \bibinfo {pages} {1760} (\bibinfo {year}
  {2012})}\BibitemShut {NoStop}%
\bibitem [{\citenamefont {Werbos}(1990)}]{werbos:backpropagation}%
  \BibitemOpen
  \bibfield  {author} {\bibinfo {author} {\bibfnamefont {P.}~\bibnamefont
  {Werbos}},\ }\href {https://doi.org/10.1109/5.58337} {\bibfield  {journal}
  {\bibinfo  {journal} {Proceedings of the IEEE}\ }\textbf {\bibinfo {volume}
  {78}},\ \bibinfo {pages} {1550} (\bibinfo {year} {1990})}\BibitemShut
  {NoStop}%
\bibitem [{\citenamefont {Hochreiter}\ and\ \citenamefont
  {Schmidhuber}(1997)}]{lstm:hochreiter}%
  \BibitemOpen
  \bibfield  {author} {\bibinfo {author} {\bibfnamefont {S.}~\bibnamefont
  {Hochreiter}}\ and\ \bibinfo {author} {\bibfnamefont {J.}~\bibnamefont
  {Schmidhuber}},\ }\href {https://doi.org/10.1162/neco.1997.9.8.1735}
  {\bibfield  {journal} {\bibinfo  {journal} {Neural Computation}\ }\textbf
  {\bibinfo {volume} {9}},\ \bibinfo {pages} {1735} (\bibinfo {year}
  {1997})}\BibitemShut {NoStop}%
\bibitem [{\citenamefont {Koolstra}\ \emph {et~al.}(2022)\citenamefont
  {Koolstra}, \citenamefont {Stevenson}, \citenamefont {Barzili}, \citenamefont
  {Burns}, \citenamefont {Siva}, \citenamefont {Greenfield}, \citenamefont
  {Livingston}, \citenamefont {Hashim}, \citenamefont {Naik}, \citenamefont
  {Kreikebaum}, \citenamefont {O'Brien}, \citenamefont {Santiago},
  \citenamefont {Dressel},\ and\ \citenamefont {Siddiqi}}]{koolstra:lstm}%
  \BibitemOpen
  \bibfield  {author} {\bibinfo {author} {\bibfnamefont {G.}~\bibnamefont
  {Koolstra}}, \bibinfo {author} {\bibfnamefont {N.}~\bibnamefont {Stevenson}},
  \bibinfo {author} {\bibfnamefont {S.}~\bibnamefont {Barzili}}, \bibinfo
  {author} {\bibfnamefont {L.}~\bibnamefont {Burns}}, \bibinfo {author}
  {\bibfnamefont {K.}~\bibnamefont {Siva}}, \bibinfo {author} {\bibfnamefont
  {S.}~\bibnamefont {Greenfield}}, \bibinfo {author} {\bibfnamefont
  {W.}~\bibnamefont {Livingston}}, \bibinfo {author} {\bibfnamefont
  {A.}~\bibnamefont {Hashim}}, \bibinfo {author} {\bibfnamefont {R.~K.}\
  \bibnamefont {Naik}}, \bibinfo {author} {\bibfnamefont {J.~M.}\ \bibnamefont
  {Kreikebaum}}, \bibinfo {author} {\bibfnamefont {K.~P.}\ \bibnamefont
  {O'Brien}}, \bibinfo {author} {\bibfnamefont {D.~I.}\ \bibnamefont
  {Santiago}}, \bibinfo {author} {\bibfnamefont {J.}~\bibnamefont {Dressel}},\
  and\ \bibinfo {author} {\bibfnamefont {I.}~\bibnamefont {Siddiqi}},\ }\href
  {https://doi.org/10.1103/PhysRevX.12.031017} {\bibfield  {journal} {\bibinfo
  {journal} {Phys. Rev. X}\ }\textbf {\bibinfo {volume} {12}},\ \bibinfo
  {pages} {031017} (\bibinfo {year} {2022})}\BibitemShut {NoStop}%
\bibitem [{\citenamefont {Farzad}\ \emph {et~al.}(2019)\citenamefont {Farzad},
  \citenamefont {Mashayekhi},\ and\ \citenamefont
  {Hassanpour}}]{farzad:lstm-activations}%
  \BibitemOpen
  \bibfield  {author} {\bibinfo {author} {\bibfnamefont {A.}~\bibnamefont
  {Farzad}}, \bibinfo {author} {\bibfnamefont {H.}~\bibnamefont {Mashayekhi}},\
  and\ \bibinfo {author} {\bibfnamefont {H.}~\bibnamefont {Hassanpour}},\
  }\href {https://doi.org/10.1007/s00521-017-3210-6} {\bibfield  {journal}
  {\bibinfo  {journal} {Neural Computing and Applications}\ }\textbf {\bibinfo
  {volume} {31}},\ \bibinfo {pages} {2507} (\bibinfo {year}
  {2019})}\BibitemShut {NoStop}%
\bibitem [{\citenamefont {Kingma}\ and\ \citenamefont
  {Ba}(2017)}]{kingma:adam}%
  \BibitemOpen
  \bibfield  {author} {\bibinfo {author} {\bibfnamefont {D.~P.}\ \bibnamefont
  {Kingma}}\ and\ \bibinfo {author} {\bibfnamefont {J.}~\bibnamefont {Ba}},\
  }\href {http://arxiv.org/abs/1412.6980} {\bibinfo {title} {Adam: {A} {Method}
  for {Stochastic} {Optimization}}} (\bibinfo {year} {2017}),\ \bibinfo {note}
  {arXiv:1412.6980 [cs]}\BibitemShut {NoStop}%
\bibitem [{\citenamefont {Chollet}\ \emph {et~al.}(2015)\citenamefont {Chollet}
  \emph {et~al.}}]{keras:chollet2015}%
  \BibitemOpen
  \bibfield  {author} {\bibinfo {author} {\bibfnamefont {F.}~\bibnamefont
  {Chollet}} \emph {et~al.},\ }\href@noop {} {\bibinfo {title} {Keras}},\
  \bibinfo {howpublished} {\url{https://keras.io}} (\bibinfo {year}
  {2015})\BibitemShut {NoStop}%
\bibitem [{\citenamefont {Montangero}(2018)}]{montangero_introduction_2018}%
  \BibitemOpen
  \bibfield  {author} {\bibinfo {author} {\bibfnamefont {S.}~\bibnamefont
  {Montangero}},\ }\href {https://doi.org/10.1007/978-3-030-01409-4} {\emph
  {\bibinfo {title} {Introduction to {Tensor} {Network} {Methods}: {Numerical}
  simulations of low-dimensional many-body quantum systems}}}\ (\bibinfo
  {publisher} {Springer International Publishing},\ \bibinfo {address} {Cham},\
  \bibinfo {year} {2018})\BibitemShut {NoStop}%
\bibitem [{\citenamefont {Skolik}\ \emph {et~al.}(2022)\citenamefont {Skolik},
  \citenamefont {Cattelan}, \citenamefont {Yarkoni}, \citenamefont {Bäck},\
  and\ \citenamefont {Dunjko}}]{skolik:geometric-equivariant}%
  \BibitemOpen
  \bibfield  {author} {\bibinfo {author} {\bibfnamefont {A.}~\bibnamefont
  {Skolik}}, \bibinfo {author} {\bibfnamefont {M.}~\bibnamefont {Cattelan}},
  \bibinfo {author} {\bibfnamefont {S.}~\bibnamefont {Yarkoni}}, \bibinfo
  {author} {\bibfnamefont {T.}~\bibnamefont {Bäck}},\ and\ \bibinfo {author}
  {\bibfnamefont {V.}~\bibnamefont {Dunjko}},\ }\href
  {http://arxiv.org/abs/2205.06109} {\bibinfo {title} {Equivariant quantum
  circuits for learning on weighted graphs}} (\bibinfo {year} {2022}),\
  \bibinfo {note} {arXiv:2205.06109 [quant-ph]}\BibitemShut {NoStop}%
\bibitem [{\citenamefont {Wu}\ \emph {et~al.}(2021)\citenamefont {Wu},
  \citenamefont {Pan}, \citenamefont {Chen}, \citenamefont {Long},
  \citenamefont {Zhang},\ and\ \citenamefont {Yu}}]{wu:gnn}%
  \BibitemOpen
  \bibfield  {author} {\bibinfo {author} {\bibfnamefont {Z.}~\bibnamefont
  {Wu}}, \bibinfo {author} {\bibfnamefont {S.}~\bibnamefont {Pan}}, \bibinfo
  {author} {\bibfnamefont {F.}~\bibnamefont {Chen}}, \bibinfo {author}
  {\bibfnamefont {G.}~\bibnamefont {Long}}, \bibinfo {author} {\bibfnamefont
  {C.}~\bibnamefont {Zhang}},\ and\ \bibinfo {author} {\bibfnamefont {P.~S.}\
  \bibnamefont {Yu}},\ }\href {https://doi.org/10.1109/TNNLS.2020.2978386}
  {\bibfield  {journal} {\bibinfo  {journal} {IEEE Transactions on Neural
  Networks and Learning Systems}\ }\textbf {\bibinfo {volume} {32}},\ \bibinfo
  {pages} {4} (\bibinfo {year} {2021})}\BibitemShut {NoStop}%
\end{thebibliography}%
\end{document}